\documentclass[%
 aip,
 amsmath,amssymb,
reprint,%
 author-year,%
 author-numerical,%
]{revtex4-1}

\usepackage{graphicx}
\usepackage{dcolumn}
\usepackage{bm}
\usepackage[mathlines]{lineno}

\usepackage[utf8]{inputenc}
\usepackage[T1]{fontenc}
\usepackage{mathptmx}
\usepackage{etoolbox}
\usepackage{units}
\usepackage{booktabs}
\usepackage{comment}
\usepackage{afterpage}
\usepackage{xcolor}
\usepackage{placeins}
\usepackage{hyperref}

\DeclareMathOperator{\Tr}{tr}
\newcommand{\pd}[2]{\dfrac{\partial #1}{\partial #2}}

\newcommand{\bd}[1]{\mathbf{#1}}
\newcommand{\ub}{\mathbf{u}}

\newcommand{\sbtt}[2]{{#1}_{\texttt{#2}}}

\newcommand{\xb}{\mathbf{x}}

\makeatletter
\def\@email#1#2{%
 \endgroup
 \patchcmd{\titleblock@produce}
  {\frontmatter@RRAPformat}
  {\frontmatter@RRAPformat{\produce@RRAP{*#1\href{mailto:#2}{#2}}}\frontmatter@RRAPformat}
  {}{}
}%
\makeatother
\begin{document}

\preprint{AIP/123-QED}

\title[The following article has been submitted to AIP Physics of Fluids. After it is published, it will be found at \href{https://publishing.aip.org/resources/librarians/products/journals}{Link.}]{High-resolution large-eddy simulation of indoor turbulence and its effect on airborne transmission of respiratory pathogens; model validation and infection probability analysis}

\author{Mikko Auvinen}
\email{mikko.auvinen@fmi.fi}

\author{Joel Kuula}
\author{Tiia Gr{\"o}nholm}
\affiliation{ Finnish Meteorological Institute, Erik Palmenin aukio 1, 00560, Helsinki, Finland
}%
\author{Matthias S\"{u}hring}
\affiliation{ Leibniz University Hannover, Institute of Meteorology and Climatology, Herrenh\"{a}user Strasse 2, 30419 Hannover, Germany
}%
\author{Antti Hellsten}%
\affiliation{ Finnish Meteorological Institute, Erik Palmenin aukio 1, 00560, Helsinki, Finland
}%


\date{\today}

\begin{abstract}
High-resolution large-eddy simulation (LES) is exploited to study indoor air turbulence and its effect on the dispersion of respiratory virus-laden aerosols and subsequent transmission risks. The methodology is applied to assess two dissimilar approaches to reduce transmission risks: a strategy to augment the indoor ventilation capacity with portable air purifiers and a strategy to utilize partitioning by exploiting portable space dividers. To substantiate the physical relevance of the LES model, a set of experimental aerosol concentration measurements are carried out, and their results are used for validating the LES model results.
The obtained LES dispersion results are subjected to pathogen exposure and infection probability analysis. Wells-Riley probability model is extended to rely on realistic time- and space-dependent concentration fields to yield time- and space-dependent infection probability fields. The use of air purifiers leads to greater reduction in absolute risks compared to the analytical Wells-Riley model, which fails to predict the original risk level. However, the two models do agree on the relative risk reduction. The spatial partitioning strategy is demonstrated to have an undesirable effect when employed without other measures. The partitioning approach may yield positive results when employed together with targeted air purifier units.
The LES-based results are examined in juxtaposition with the classical Wells-Riley model, which is shown to significantly underestimate the infection probability, highlighting the importance of employing accurate indoor turbulence modeling when evaluating different risk-reduction strategies.
\end{abstract}

\maketitle

\section{Introduction} \label{sec:intro}

The global COVID-19 pandemic, caused by SARS-CoV-2 virus and its variants, brought about a debate concerning the relevant transmission modes of the disease. The initially adopted view clung to the paradigm that most probable infection route involves large respiratory droplets which are ballistically sprayed due to respiratory activity either directly into the eyes, nose or mouth of a bystander or onto surfaces from which they are subsequently carried to the body openings with mucous membranes. Thus, keeping a 'safe distance' (1-2~m) to others and practicing hand hygiene were identified as sufficient preventive measures against transmission. This picture was brought into question early on as evidence began to accumulate on cases where the only viable transmission mechanisms pointed to airborne transmission or more specifically transmission by inhalation of non-ballistic virus-laden aerosol particles originating from the respiratory activity of the infected person. Despite the prolonged institutional reluctance to accept the relevance of airborne transmission mode, the general consensus has shifted under mounting evidence recognizing now that airborne transmission is the primary mode fueling the global pandemic \cite{Zhang2020}.

It is recognized that the respiratory droplets, exhaled by an infected individual, are generated at a distribution of sizes ranging from smaller than $\unit[1]{\mu m}$ up to larger than $\unit[100]{\mu m}$ in diameter, see \citet{Vuorinen2020} and references therein, e.g. \citet{Lindsley2013, Bake2019, Lindsley2012}, and \citet{Katre2021}. The largest do follow ballistic trajectories landing quickly on nearby surfaces whereas the smaller droplets undergo evaporation within seconds forming droplet nuclei which act as infective aerosol particles suspended in air \citep{Lindsley2013,Vuorinen2020}. 

The concentration of remaining small particles disperses into the surrounding air in accordance with the prevailing flow conditions, like any neutrally buoyant gas release, without significant deposition\citep{Vuorinen2020,Hinds1999}. Thus, in an indoor environment, the evolution of respiratory aerosol concentration field is solely dictated by the indoor flow system which is driven by ventilation inlet and outlet conditions and thermal gradients. 
This study brings to focus the essential turbulent mechanisms affecting the longer-range transport of infective airborne pathogens, which have played a central role in numerous super-spreading events that have been documented during the COVID-19 pandemic \cite{Shen2021}.
  
Indoor air flows are challenging to model by means of Computational Fluid Dynamics (CFD). The flow systems are complex combinations of different turbulent source mechanisms and the flow domains are largely occupied by weak free turbulence which thereby becomes the primary agent responsible for the dispersion outcomes of respiratory aerosol emissions. Such flow systems are highly sensitive to modeling errors and notoriously ill suited for Reynolds Averaged Navier-Stokes (RANS) approaches where all the turbulent motion and related effects are modeled instead of resolved. The computationally less demanding RANS approach only provides the mean flow field and mean kinetic energy of the fluctuations. Turbulence resolving techniques, such as Large-Eddy Simulation (LES), offer a more reliable approach to model indoor flow and dispersion problems because it is able to provide more direct information about the turbulent fluctuations which carry out the dispersion work.  The challenges in LES modeling, in turn, stem from the difficulty in choosing sufficient spatial resolution, which dictates the division between the resolved and modeled turbulent scales, and settling on the acceptable computational cost for the simulations. Furthermore, as the LES results for dispersion events are always individual realizations which involve inherent turbulent randomness, the statistical representativeness of the obtained results must be taken into account and addressed appropriately. 

In this study, high-resolution LES modeling with modern super-computing capacity is exploited to examine a realistic indoor ventilation flow system which features the level of complexity found in spaces used for social gatherings. Although the study employs a specific spatial and ventilation configuration, the objective is to gain physically sound insight also on the nature of mechanically ventilated indoor flows in general. The study also seeks to clarify and strengthen the understanding concerning the role air purifiers play in mechanically ventilated spaces where they augment the existing ventilation capacity. The emphasis in this study is placed on recognizing the physical mechanisms and their relevance in the context of indoor air hygiene and demonstrating a reliable LES methodology to evaluate different air hygiene strategies. This study employs a rigorous adaptation of Wells-Riley infection-probability analysis to cast the obtained LES results to relevant context. The topic of air filtration in the context of airborne transmission of SARS-CoV-2 has been actively discussed in literature (examples include \citep{Morawska2020,Lindsley2021,Heo2021,REHVA2021,Talib2021}), but the modeling approaches have systematically demonstrated numerical or other modeling deficiencies which have not been critically examined. Several LES-based indoor-dispersion studies have been published recently, e.g. \citet{Liu2021, Foster2021} to give just two examples. A number of others have been published recently but, again, their LES practices and particularly the numerical resolution choices demonstrated systematic inadequacies and are thus not necessarily comparable with the present study. 

The predictive capacity and physical validity of any numerical modeling approach must be established before the obtained results can be used to justify actions and guide decisions. Due to the absence of indoor-specific dispersion datasets, which would encompass the complexity of real mechanical ventilated flow systems, this work documents a transparent validation study to assess the LES model's performance in the context of aerosol dispersion under realistic indoor conditions. The experimental setup replicates a staged restaurant scenario where a single infected individual attends a dinner, or other social gathering, where people remain seated and exposed to each other via air for a prolonged duration. For the dispersion process examined in this study, the relevant time scale is determined to be approximately one hour. The experiment involves particulate matter sensors in 27 points, distributed evenly throughout the test room, recording the evolution of aerosol concentration originating from the infected individual. The validation study involves two ventilation configurations, the first featuring a generic, baseline ventilation and the second an augmented ventilation configuration with added air purifiers. The level of agreement between the experimental and LES modelled concentration time series is established with a group of performance metrics adopting a strict set of evaluation criteria used in air-quality modeling\citep{Chang2004,Moonen2013}.  

Having demonstrated that the LES model is able to resolve the principal flow physics governing the evolution of respiratory aerosol concentration fields indoors, the analysis can be broadened to consider the exposure of other subjects to the viral load emitted by the infected host and, using well-justified models, determine their probability to contract an infection.  

The widely applied analytical Wells-Riley infection-probability model, based on the early works by \citet{Wells1955} and \citet{Riley1978} and its later extension for time-dependent concentration by \citet{Gammaitoni1997}, is founded on the assumption of perfect mixing which occurs instantaneously, resulting in a spatially constant concentration of pathogen-laden aerosols indoors. 
(Note that nowadays the name Wells-Riley model usually refers to the extended version by \citet{Gammaitoni1997}.) The assumption of spatially constant concentration has the advantage that the concentration is an analytical solution of a simple linear ordinary differential equation. 
Exploiting this result in another linear ordinary differential equation yields a solution for infection probability that is also time dependent, but spatially uniform everywhere. 
However, spatially constant concentration is a strong simplification, since, in reality, the concentration field varies strongly as a function of indoor turbulence and ventilation conditions. We hypothesize that this systematically leads to strong underestimation of the mean concentration, and subsequently resulting in over-optimistic infection probabilities which is an unacceptable trait for a risk-assessment tool. Situations where the aerosol source (host) are in the immediate proximity of an outlet vent or an air purifier do constitute exceptions, but the probability of such circumstances is judged insignificant. Moreover, the constant-concentration assumption ignores peak concentrations and thus peak probabilities leading to further underestimation. The study by \citet{Zhang2021} already provided preliminary indication of these shortcomings. They carried out a case study focusing to the interior of a short-haul bus in high and low ventilation settings. They compared unsteady RANS (URANS) approach with the analytical Wells-Riley model. They concluded that the Wells-Riley results differ from the CFD-results, not only locally due to its inability to capture any spatial variability, but also in the spatially averaged sense, especially when the ventilation rate is small.

Realistic approximations for time- and space-dependent concentration fields can only be obtained by carefully implemented high-resolution CFD modeling. Combining CFD-data with the Wells-Riley probability model has been suggested in several studies \cite{Qian2009, Gupta2012, Yan2017, You2019, Guo2021, Foster2021}. However, none of these studies include mathematical or statistical justification for substituting spatially variable concentration data into the Wells-Riley probability model which is based on the assumption of spatially constant concentration. In this study, the Wells-Riley probability model is formally and rigorously extended to rely on CFD-predicted time- and space-dependent realistic concentration fields and to yield time- and space-dependent realistic infection probability fields. The results are compared with those from the original model \cite{Gammaitoni1997}. 

A number of recent studies have focused on the infection risk due to instantaneous emission events such as coughs or sneezes, e.g. \citet{Vuorinen2020, Balachandar2020, Liu_Balachandar2021}. In contrast, this study focuses on longer-term exposure from continuous emission period which can be conceived to consist of various respiratory activities (breathing, talking, laughing, etc). In addition to studying the infection risk in the baseline set up of the case study, two risk-reduction strategies are assessed. The first strategy is to use air-purifiers, which in this study augment the existing baseline ventilation rate by 65\%. The second strategy is somewhat controversial exploiting partitioning where the dining tables are separated from each other using $\unit[1.5]{m}$ screens as space dividers. The influence of applying both concurrently is also examined. At the beginning of this investigation, it was reasonable to expect that air purifiers can reduce the infection risk (in some accordance with their relative filtration capacity), but the expectation for the effectiveness of space dividers was adverse. However, partitioning strategy was included in this study because it has been observed that there is wide-spread popular belief concerning its effectiveness.  

The article is organized as follows. Section~\ref{sec:setup} describes the restaurant space and the study setup. Section~\ref{sec:exp_methods} documents the experimental methods and Section~\ref{sec:num_methods} the numerical modeling, including the modeling methods and the validation approach. Section~\ref{sec:results} presents the numerical model validation, turbulence results and infection probability analysis. This section also reports the comparison between the selected risk-reduction strategies. The extension of the infection probability model to spatially variable concentration fields is presented in Section~\ref{sec:results} albeit its analytical details are given in Appendix~\ref{app:P_extension}. Finally, the conclusions are drawn in Section~\ref{sec:conclusions}.

\section{Description of study setup} \label{sec:setup}

\begin{figure*}[ht]
    \centering
    \includegraphics[width=0.85\textwidth]{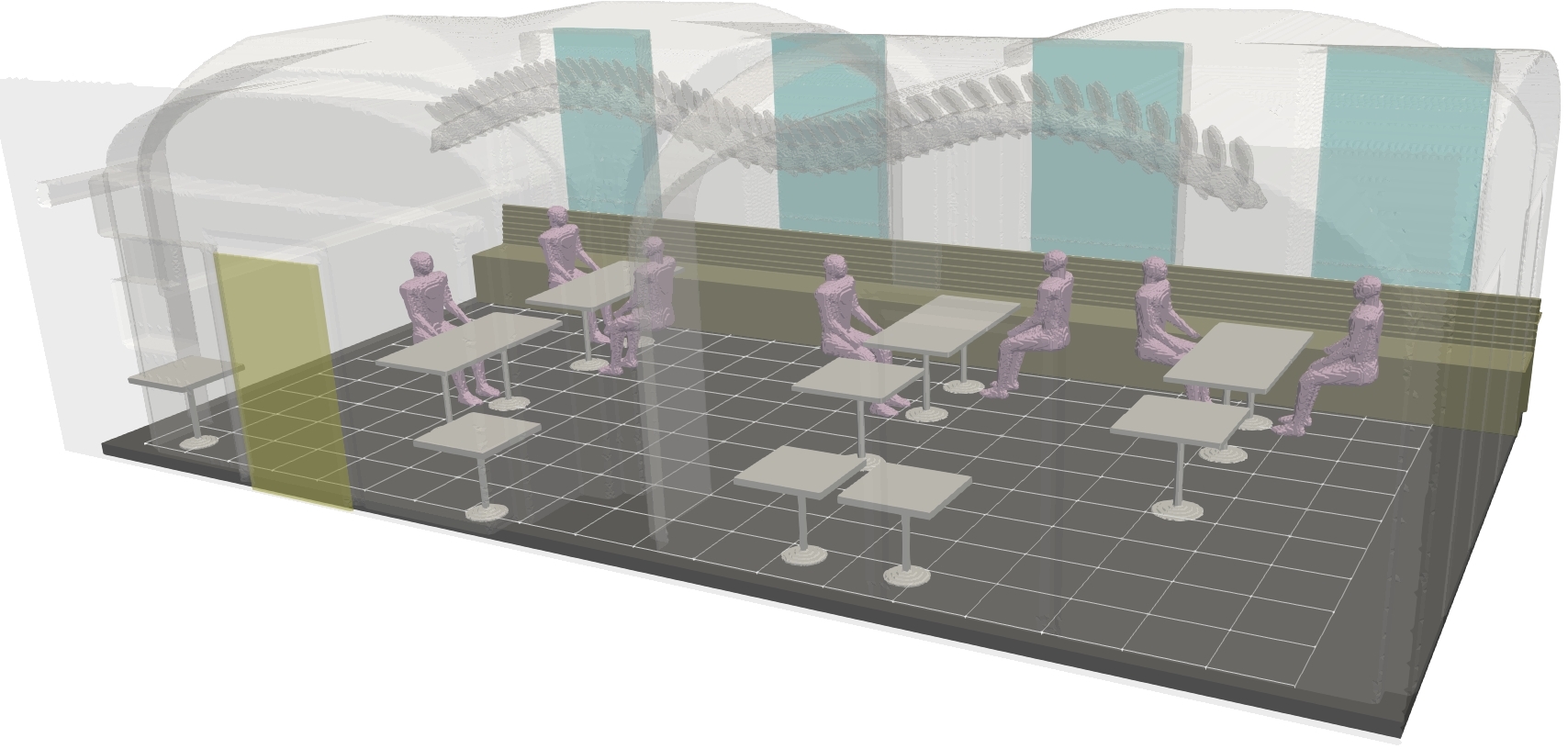}
    \caption{Visualization of the restaurant room facilitating the study. Four windows on the rear wall are coloured with cyan and the entrance opening with yellow. The entrance is sealed shut for the experiments. The wavy-shaped chain-like overhanging structure is a sculpture. A grid with $\unit[0.5]{m}$ spacing is shown on the floor for visual assistance.}
    \label{fig:room_overview}
\end{figure*}

The study exploits a real restaurant room located downtown Helsinki, Finland, as a laboratory for the experiments and modeling. The aerosol dispersion experiments were conducted onsite while the computational LES model was constructed in accordance to detailed laser-scanning data sets obtained from the interior of the restaurant room. The room's floor area is approximately  $\unit[60]{m^2}$ and its effective interior volume approximately $\unit[170]{m^3}$. A computer generated visualization of the geometric layout of the interior is shown in Fig.~\ref{fig:room_overview} featuring a generic table and seating arrangement employed in this study. 

\subsection{Ventilation and air filtration} \label{sec:vent}

The restaurant building is from the early 1900's, but it has been augmented with a modern centrally operated mechanical ventilation system (forced flow in both inlets and outlets) such that the room features three inlets feeding fresh, heated air toward the windows and three outlets on the opposite wall from the windows. Under normal circumstances the room's outlet ventilation capacity is shared by an adjacent room connected via an entrance opening (shown in yellow in Fig.~\ref{fig:room_overview}. As the second room also features an inlet, the outlet capacity of the lab room becomes excessive when it is sealed for the aerosol experiments. To reduce the resulting imbalance between the room's inlet and outlet flow rates, one of the outlets is blocked. Further details on the inlet and outlet duct locations and their respective ventilation rates are provided in Section~\ref{sec:num_bc_grid}.

The room and its ventilation flow system exhibits all the relevant complexities characterizing real indoor environments having highly variable flow and temperature distributions arising from mechanical ventilation (inlet jets, outlet suction), heating elements (radiators, human bodies) and cooling surfaces (windows during cold seasons). The complexity ensures that the studied indoor flow system is representative of a range of realistic mechanical indoor ventilation scenarios where the flow is driven by a multitude of turbulence generating mechanisms.  

In general, the most notable deviation separating real and modelled indoor flow systems stem from the natural imperfections found in actual buildings. Buildings, particularly older ones, are not perfectly sealed from the ambient (in contrast to the computational models) but leak through windows and other seams and cracks in the structures. Such leaks do manifest in the volume flow rate budget of this study's lab room. The forced ventilation maintains the room at low pressure in relation to the ambient resulting in leakage flows at windows and other small gaps in the structures giving rise to a flow rate imbalance between the inlet and outlet ducts. This non-idealization is accepted and addressed in the context of model evaluation (see Section~\ref{sec:model_evalution}). 

The effect of added indoor air filtration is examined by considering nominal ventilation scenarios in juxtaposition with situations where the room is equipped with two portable UniqAir PRO \cite{UniqAir} air purifiers. This study employs air purifiers as a potential means to reduce transmission risks indoors by introducing locally administered filtration and increasing turbulent mixing within the room. However, this study does not examine the criteria for sufficient risk reduction. Such examinations which categorize different risk levels, fall beyond the scope of this work which focuses on the flow physics governing airborne transmission events. The air purifiers used in this study employ a two-stage filtration featuring a HEPA filter and an active carbon filter in series with a reported $>99\%$ separation rate for particles $\unit[>0.1]{\mu m}$ in size. The device has dimensions $\unit[31.7]{cm}$, $\unit[31.7]{cm}$ and $\unit[110.4]{cm}$ in width, depth and height respectively. The maximum power consumption is $\unit[85]{W}$ per device yielding a $\unit[330]{m^3}$ per hour flow rate. Details on how air purifiers were operated and modelled in this study are laid out in Section~\ref{sec:numerical_vent_filt}.

\section{Experimental methods}  \label{sec:exp_methods}

\subsection{Aerosolization method}
\label{sec:exp_nebu}
\begin{figure}[b!]
\includegraphics[width=78mm]{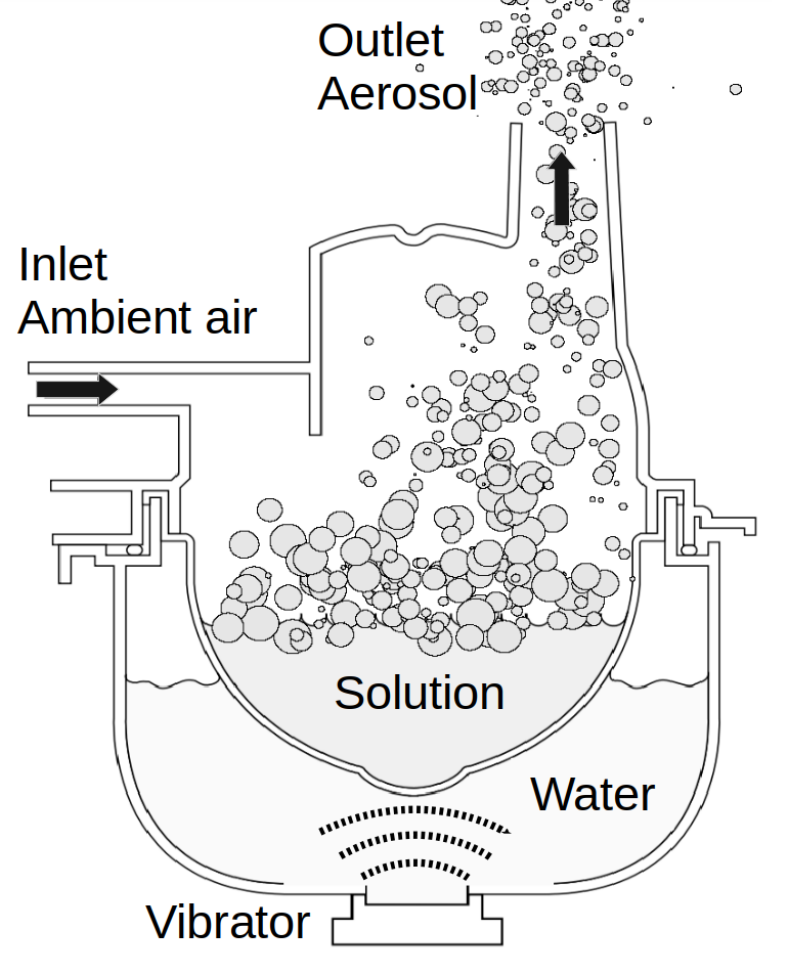}
\caption{Schematic illustration of the ultrasonic nebulizer filled with potassium phosphate solution. Ultrasonic vibration is transmitted through water into the solution. Due to vibration, aerosol particles are released and carried out from the chamber by air flow. The wet size of the produced particles were 6-8 $\mu$m.} \label{fig:nebu}
\end{figure}

The aerosol particles were generated by Omron Ultrasonic Nebulizer Model NE-U17. It consists of an ultrasonic vibrator (frequency $\unit[1.7]{MHz}$) which is located at the bottom of the water tank. In the tank we used $\unit[375]{mL}$ of ultra-pure water (Milli-Q). Above the water tank, there is a aerosolization chamber for the solution which will be converted into aerosol in the nebulization process. The device includes a fan which causes the filtered ambient air to flow through the upper part of the aerosolization chamber and a $\unit[70]{cm}$ long plastic hose with $\unit[20]{mm}$ diameter. The vibration is transmitted through water into the aerosolization chamber. The energy of the ultrasonic vibration forms aerosol on the surface of the solution. Aerosol is carried out from the device by the air flow. The nebulization arrangement is illustrated in Fig.~\ref{fig:nebu}. In the beginning of each simulation we filled the aerosolization chamber with 150~ml of $\unit[20]{mmol/L}$ potassium phosphate ($\mathrm{K}_3\mathrm{PO}_4$) in $\unit[1]{mmol/L}$ $\mathrm{MgCl}_2$ solution. The nebulization rate was approximately $\unit[0.3]{mL \, min^{-1}}$ and according to our measurements the volume flow rate was approximately $\unit[14]{L \, min^{-1}}$. Based on our laboratory tests, Omron nebulizer produced $\unit[6-8]{\mu\mathrm{m}}$ sized wet particles, dry size being $<\unit[1]{\mu m}$ in diameter from the solution used in experiments. The mode of the dry size distribution was approximately $\unit[0.9]{\mu m}$. By assuming uniform distribution of generated particles $\unit[7]{\mu m}$ in diameter, the nebulization rate was approximately $3\cdot10^8$ particles per second.

Due to the evaporative cooling taking place immediately after the aerosol exits the nebulizer outlet, the plume becomes negatively buoyant and it tends to sink right after the outlet. Naturally, the largest particles tend to sink even in still air, but the evaporative cooling effect makes the air itself sink due to the negative buoyancy and the sinking of air moves even the smallest particles downwards. The downdraft close to the outlet is an undesired phenomenon since the aim was to mimic human exhaling which is positively buoyant. To alleviate this discrepancy, a thin heated stainless steel plate (dimensions  $\unit[44 \times 59]{cm}$) with a $\unit[20]{W}$ heating cable installed below the plate was placed below the outlet of the nebulizer to avoid immediate sink of the air. The surface temperature of the plate was approximately $\unit[20]{^\circ C}$ during the nebulization largely eliminating the evaporative cooling effect. The air temperature above the plate edge opposite to the outlet was measured to be $\unit[20-21]{^\circ C}$ a while the ambient room air temperature was roughly $\unit[19]{^\circ C}$. The air temperature in the aerosolization chamber near the outlet hose joint reached $\unit[31-33]{^\circ C}$ towards the end of the simulation.

\subsection{Measurement devices and their arrangement}
To analyze the validity of the LES-model, real-time aerosol particle dispersion measurements were performed on Feb 1st, 2021 using two model 3321 Aerodynamic Particle Sizers (APS1 and APS2, TSI Inc., USA) and nine SPS30 particulate matter (PM) sensors (Sensirion AG, Switzerland). The reference instrument APS is a time-of-flight-based particle size spectrometer, which measures the number and aerodynamic size of particles from 0.5 to 20 $\mu$m with a 52-bin resolution (\citet{Peters2003}). The SPS30 sensor, on the other hand, is an optical low-cost sensor, which measures PM number concentrations from approximately 0.3 to 10 $\mu$m with a 5-bin resolution. Previous laboratory evaluation has shown, however, that the SPS30 does not precisely adhere to its declared technical specifications, and it is in fact best suited for the measurement of particles smaller than 1.3 $\mu$m (\citet{Kuula2020}). The particle size distribution produced in the nebulizer had a mode of $\unit[< 1]{\mu\mathrm{m}}$ and therefore the limited detection range of the SPS30 was not an issue; only the two first size bins of the SPS30 were used in this study. The nine SPS30 sensor nodes were equipped with Sensirion SHT85 temperature and humidity sensors as well. This allowed for the monitoring of temperature and relative humidity gradients within the study room. To ensure data uniformity across all devices, the SPS30 sensors as well as the APS2 were calibrated using the APS1 as a reference. The calibration was conducted by co-locating the devices and then performing measurements over a concentration range of approximately $\unit[10-2000 ]{cm^{-3}}$. Similar to this, the SHT85 environmental sensors were calibrated to ensure unit to unit uniformity. The air flow rates in the room ventilation ducts (both in supply and outgoing air ducts) was measured with VelociCalc Air Velocity Meter, Model 8347 (TSI Inc., USA) before starting the experiment. 

\begin{figure*}[ht]
    \centering
    \includegraphics[width=0.8\textwidth]{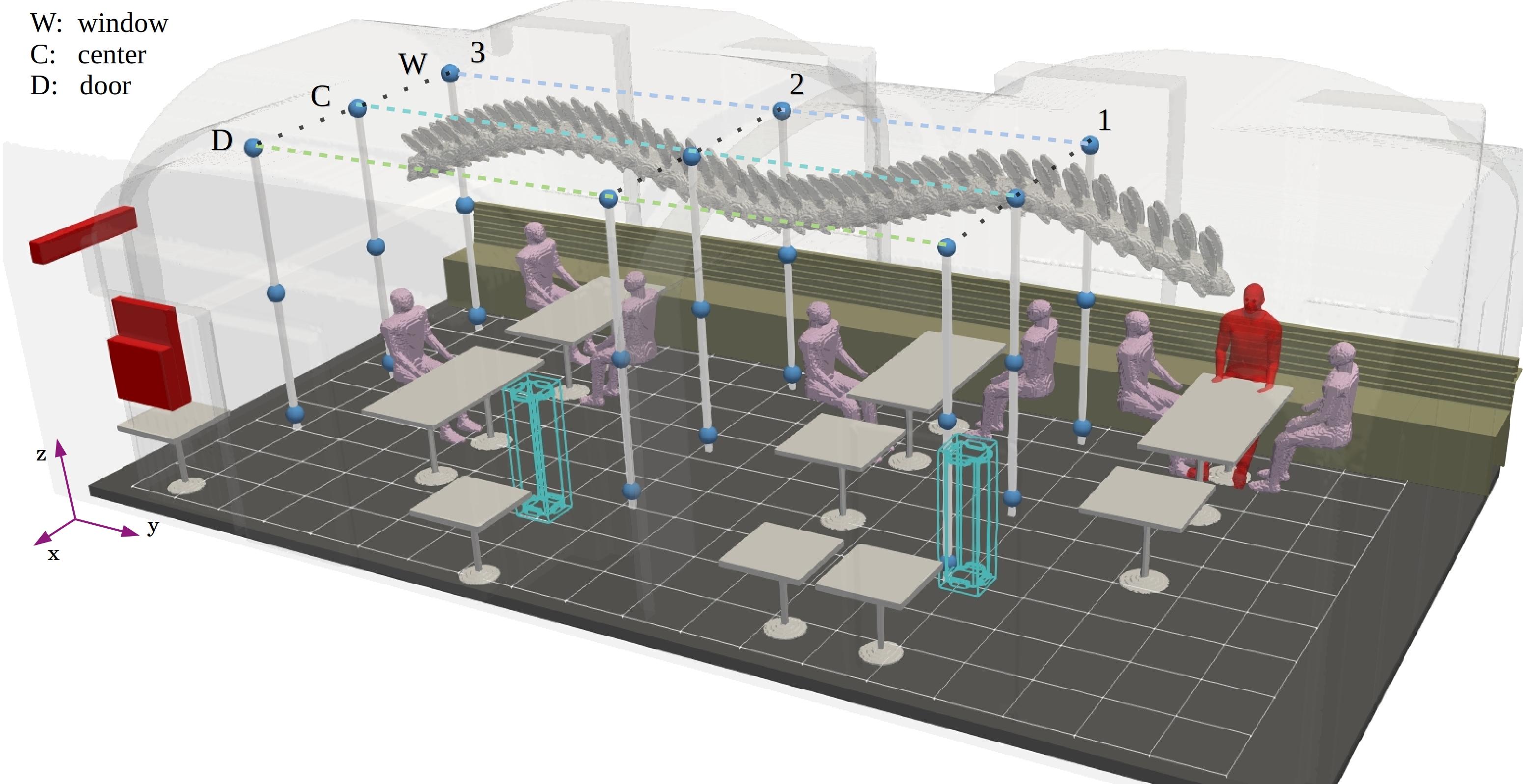}
    \caption{Overview of the sensor arrangement and their naming convention. The sensor coordinates are labelled such that W, C and D indicate the location in the $x$-direction as shown, numbers 1, 2 or 3 specify the location in the $y$-direction (1 being closest to the aerosol source) and vertical positioning is given by LO, MID and HI in accordance with the sensor height. The aerosol source is situated to coincide with the mouth of the imaginary infected individual who is shown in red at the end of the table closest to the W1 sensor mast.}
    \label{fig:sensor_layout}
\end{figure*}

An illustration of the measurement arrangement is shown in Figure~\ref{fig:sensor_layout}. The nine sensor nodes were attached to three separate masts (three nodes per mast) at equally spaced heights from $\unit[25]{cm}$ to $\unit[210]{cm}$ (LO, MID, and HI) above floor level. Within the study room, the three masts were then spaced equally along the $y$-axis. When conducting an experiment, the row of the three masts were moved between different $x$-axis coordinates W, C, and D (Window, Center, and Door) to capture a total of 27 measurement points. The APS1 and APS2 were used to measure the particle size distributions at different locations within the room, and their location varied from one experiment to another. The measured particle number concentration data used to evaluate LES-model relied solely on the data measured with the sensor nodes.\\

\subsection{Particle dispersion experiment}

Aerosol particle dispersion experiments were conducted on a generic seating configuration with (GEN+FLT) and without (GEN) air purifiers. Both experiments were divided into three parts in order to cover all 27 aerosol particle sampling points. Each part of an experiment began by ventilating the room so that the current aerosol particle number concentration corresponded to that of the general background concentration. Then, nebulizer was switched on and the dispersion of aerosol particles was observed using the nine sensor nodes attached to the three masts. After the experiment had lasted 60 minutes, the nebulizer was switched off and the room was thoroughly ventilated. This process was carried out three times both in GEN and GEN+FLT configurations in order to cover all three mast-row positions W, C and D, see Fig.~\ref{fig:sensor_layout}. 

Contrary to Fig.~\ref{fig:sensor_layout} no humans were inside the room during the dispersion experiment. This choice was made after test computations revealed that the presence of few people as solid bodies and heat sources is insignificant for the particle dispersion in this experimental and modeling arrangement.

\section{Numerical methods} \label{sec:num_methods}

\subsection*{Note on Nomenclature}  
The nomenclature conventions used throughout this article are presented herein. All Cartesian vectors are denoted by boldface font and scalars by normal font. For instance the coordinate vector and its components are denoted as $\xb = (x, y, z)$. In the context of LES, the velocity vector complies with decomposition $\mathbf{U} = \ub(\xb,t) + \ub''(\xb,t)$ where $\mathbf{U}$ is the unfiltered velocity field, $\ub$ is the grid-resolved (space-filtered), and  $\ub''$ is the subgrid-scale velocity vector. The velocity vector components are $\ub = (u, v, w)$. When required, spatial filtering in LES is denoted by tilde $\widetilde{\,\bullet\,}$. In this article, the temporal averaging is denoted by overbar $\overline{\,\bullet\,}$ and spacial averaging by angled brackets such that volume averaging $\left<  \bullet  \right> \equiv \left< \bullet \right>_{V}$ is indicated by default while in other situations a subscript is used. All spatial averages are intrinsic averages where data points within solid obstacles are excluded. The Reynolds decomposition of the velocity field adopts the following notation: $\ub(\xb,t) = \overline{\ub}(\xb) + \ub'(\xb,t)$, using prime $(\bullet)'$ to specify fluctuations about the temporal mean.

\subsection{The PALM model system}  \label{sec:num_palm}

The PALM large-eddy simulation model system \citep{Raasch2001, Maronga2015, Maronga2020} is designed for atmospheric boundary layer modeling, but in this work it is adapted for the present indoor flow problem. Usually general-purpose CFD models are used for for this kind of problems, but we decided to use a slightly modified version of PALM because it is computationally more efficient than many general-purpose CFD-models.  In the present configuration PALM solves the filtered Navier-Stokes-equations in the Boussinesq-approximated form. 

\begin{align}
\label{eqn:continuity}  
 & \nabla \cdot \ub  = 0   \; ,      \\
\label{eqn:momentum}    
 & \pd{\ub}{t} = - \nabla \cdot ( \ub \otimes \ub ) - \frac{1}{\rho_0} \nabla \pi^{\ast} 
 + g \frac{\theta - \langle \theta \rangle_{xy}}{\langle\theta\rangle_{xy}} \bd{\hat{k}}
  + \nabla \cdot  \bd{T}  + \bm{\Phi}_{u} \\ 
\label{eqn:energy} 
& \pd{\theta}{t} = - \ub \cdot \nabla \theta - \nabla \cdot 
\bm{\Gamma}_\theta
+ \Phi_{\theta}  \\
\label{eqn:concentration} 
& \pd{c}{t} = - \ub \cdot \nabla c - \nabla \cdot 
\bm{\Gamma}_c
+ \Phi_{c} , 
\end{align}
In Eqn.~\eqref{eqn:momentum}, $g$ is the gravitational acceleration, $\bd{\hat{k}}$ the unit vector in $z$-direction, $\pi^{\ast} = p^{\ast} + 1/3 \rho_0 \Tr{\left(\sbtt{\bd{T}}{sgs}\right)}$ is the modified perturbation pressure ($p^{\ast}$ being the perturbation pressure), $\rho_0$ is the constant density of air at ground level, $\bm{\Phi}_{u}$ is the momentum source vector and $\bf{T}$ the effective stress tensor:
\begin{equation}
 \bd{T} = \bd{T}_{\nu} - \sbtt{\bd{T}}{sgs}  
  + {\frac{1}{3}} \Tr{ \left( \sbtt{\bd{T}}{sgs}\right) \bd{I} } \, .
\end{equation}
Here, $\bd{I}$ is the identity matrix,  $\bd{T}_{\nu} = \nu  \left(  \nabla \ub + \nabla \ub^{\textsc{t}} \right)$ is the viscous stress tensor $\nu$ being the kinematic viscosity of air at room temperature, $\sbtt{\bd{T}}{sgs} = \left( \widetilde{ \mathbf{U} \otimes \mathbf{U}} - \widetilde{\ub} \otimes \widetilde{\ub} \right) $ is the subgrid-scale kinematic stress tensor where $\otimes$ denotes the outer vector product.  

In the energy equation~\eqref{eqn:energy} $\theta$ is the air temperature, $\Phi_{\theta}$ is the temperature source term and $\bm{\Gamma}_\theta$ constitutes the subgrid scale temperature flux vector. Similarly in Eqn.~\eqref{eqn:concentration}, $c$ is the scalar concentration,  $\Phi_{c}$ is the concentration source term and $\bm{\Gamma}_c$ is the subgrid scale concentration flux vector.
The temperature and concentration flux vectors are
\begin{eqnarray}
\bm{\Gamma}_\theta &=& D_\theta \nabla \theta - \left( \widetilde{\mathbf{U} \Theta} - \widetilde{\ub} \widetilde{\theta} \right)  \\
\bm{\Gamma}_c &=& D_c \nabla c - \left( \widetilde{\mathbf{U} C} - \widetilde{\ub} \widetilde{c} \right) \, ,
\end{eqnarray}
where $D_\theta$ and $D_c$ are the corresponding molecular diffusion coefficients.

PALM solves the  prognostic equations on a staggered Cartesian Arakawa-C grid. Subgrid-scale turbulence is parameterized using a 1.5-order closure after \citet{Deardorff1980} and \citet{Saiki2000}. In this study PALM solves seven prognostic quantities: the resolved velocity components $\ub = (u, v, w)$, the temperature $\theta$, concentration $c$, the subgrid-scale (SGS) turbulent kinetic energy $e$ and the modified perturbation pressure $\pi^{\ast}$. 

The discretization in time and space is achieved using a third-order Runge-Kutta scheme after \citet{Williamson1980} and a fifth-order advection scheme after \citet{Wicker2002} advection terms and second-order central difference scheme for diffusion terms. The horizontal grid spacing is always equidistant, whereas it is possible to use variable grid spacing in the vertical direction albeit this feature was not utilized in this work. 

The incompressible flow solver algorithm employs a predictor-corrector approach where the predictor step involves the calculation of a preliminary prognostic velocity from Eqn.~\eqref{eqn:momentum} whereas the corrector phase imposes continuity by numerically solving a Poisson equation for modified perturbation pressure whose solution is subsequently used to make the velocity field divergence free after every Runge-Kutta sub-time step \citep[e.g.][]{Patrinos1977}. Here, an iterative multigrid scheme (see \citep[e.g.][]{Hackbusch1985}) is used to solve the Poisson equation.

Parallelization of PALM is achieved by using the Message Passing Interface \citep[MPI, e.g.][]{Gropp1999} and a two-dimensional (horizontal) domain decomposition approach.

\subsection{Indoor model, boundary conditions and discretization}  \label{sec:num_bc_grid}

\begin{figure*}[ht]
    \centering
    \includegraphics[width=0.9\textwidth]{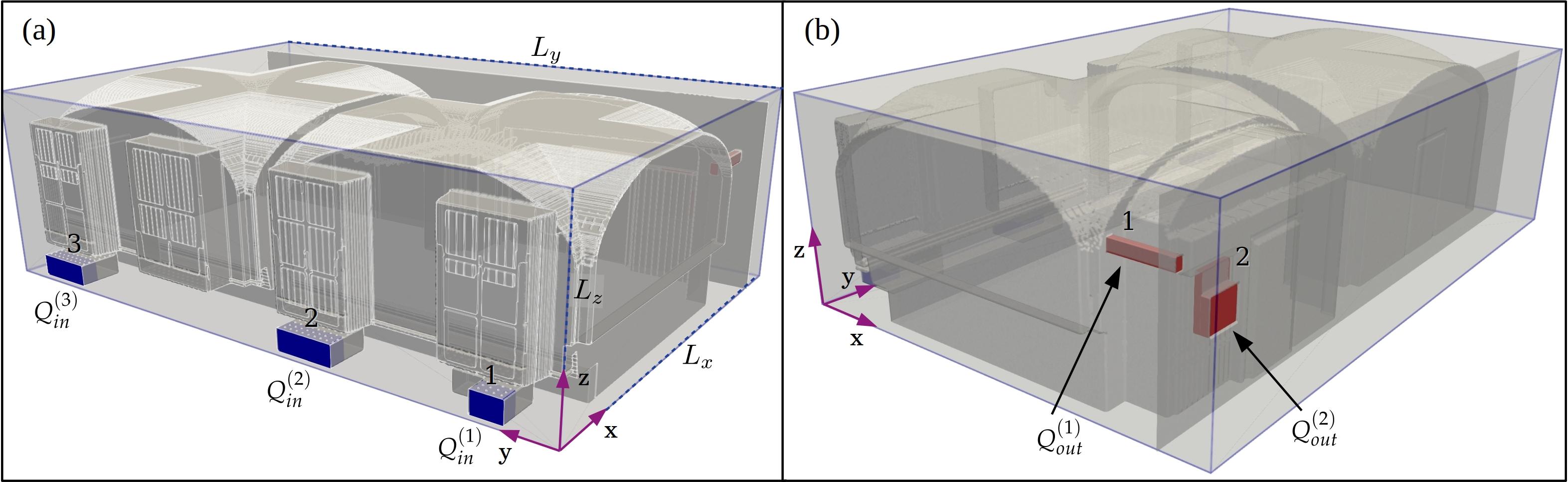}
    \caption{Front (a) and rear (b) view of the complete LES domain with inlet and outlet ducts highlighted in blue and red respectively. The inlet boundary conditions for momentum are set by imposing fixed velocity value on each inlet plane. The desired volume flow rate $Q^{(i)}_{in}$ is thus obtained by sizing the cross-sectional area $A^{(i)}$ of each inlet duct.}
    \label{fig:bc_overview}
\end{figure*}

\begin{figure*}[ht]
    \centering
    \includegraphics[width=0.9\textwidth]{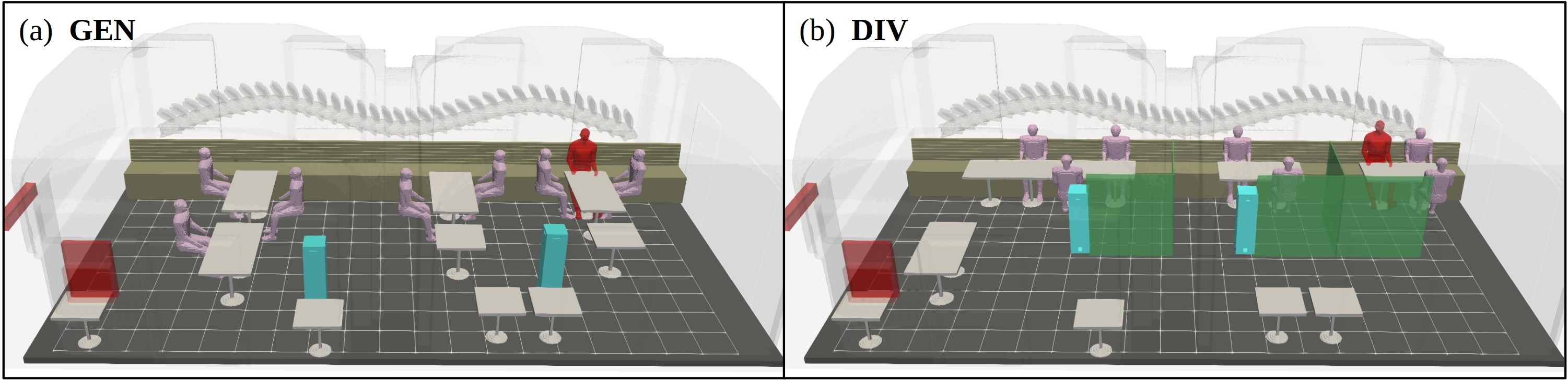}
    \caption{Overview of the LES model featuring all solid objects within the restaurant. Generic (GEN) seating configuration is displayed in (a) while an alternative configuration with $\unit[1.5]{m}$ tall space dividers (DIV) is shown in (b). A grid with $\unit[0.5]{m}$ spacing is drawn on the floor for spatial reference.}
    \label{fig:les_config}
\end{figure*}

The indoor flow model is constructed from a Cartesian grid whose dimensions and resolutions are tabulated in Table~\ref{tab:domain}.
An overview of the regular computational domain and the embedded room geometry, with highlighted ventilation inlet and outlet ducts, is visualized in Figure~\ref{fig:bc_overview}. The effective indoor air volume of the room (excluding all solid objects) is $V = 170~\unit{m^3}$. 
The solid geometries (walls, tables, sofas, humans, sculptures) are embedded within the grid by cell-masking method. The solid objects and building structures are first generated as solid 3D mesh objects with a conformal resolution and the cell-center points are then mapped onto the PALM grid with appropriate translation and rotation operations. The solid 3D meshes were generated using OpenFOAM's \textit{snappyHexMesh} tool \cite{OpenFOAM}.  Figure~\ref{fig:les_config} lays out the two different model configurations considered in this study. The generic reference (GEN) configuration features a conventional table arrangement whereas the alternative configuration employs space dividers (DIV) in effort to provide alleged shielding between customer groups. The table and seating arrangements are designed to cater the needs of the experimental work conducted at the restaurant and do not fully reflect real operational layouts used in practice. The number and placement of the humans within the room is in accordance with the experimental setup.

\begin{table}[ht]
\caption{Dimensions and grid specifications for the indoor LES domain.}
\centering
\begin{tabular}{rl}
 \toprule
  Domain dimensions:      &  $( L_x, L_y, L_z ) \; = \; ( \, 7.10 , \, 10.64 , \, 3.10 \, )$ m \\[1.5ex]
  Resolution:             &  $( \Delta x, \Delta y, \Delta z ) \; = \; ( \, 9.24, \, 9.24, \, 9.69 \, ) \times 10^{-3} \, \mathrm{m}$ \\[1.5ex]
  Domain node counts:     &  $( N_x, N_y, N_z) \; = \; ( \, 768, \, 1152, \, 320 \, )$\\[1.5ex]
 Total domain node count: &  $ N_x N_y N_z \; = \;  283 \times 10^6$  \\
\bottomrule
\end{tabular}
\label{tab:domain}
\end{table}

The momentum inlet boundary conditions are imposed by setting fixed values $u|_{\xb \in A(i)}  = U_{in}$ for every $i^{\mathrm{th}}$ inlet boundary and sizing each inlet duct area $A^{(i)}$ such that desired volume flow rates $Q^{(i)}_{in}$ are attained. The inlet and outlet ducts are highlighted in Figure~\ref{fig:bc_overview} for clarity. The inlet ducts form a U-shape and inject the fresh air into the room through a small opening facing in the negative $x$-direction. The jets are targeted straight at the lower portion of the windows. 

Due to limitations of the finite difference formulation in PALM, the mass conservation is not guaranteed in the smallest inlet and outlet duct sections and therefore the continuity is reinforced with locally applied momentum volume source terms in the relevant duct sections. This treatment guarantees $\sum_i Q^{(i)}_{in} = \sum_j Q^{(j)}_{out}$  without influencing the resolved turbulence within the room. In fact, utilizing such local momentum source terms in the ventilation ducts proved to be a highly convenient approach to control and adjust the ventilation conditions of the model. The ventilation outlet ducts discharge the indoor air into a roughly $\unit[1]{m}$ wide open volume preceding the computational domain outlet (at $x = L_x$) which acts as a conventional zero-gradient boundary for velocity terms.

\begin{figure*}[ht]
    \centering
    \includegraphics[width=0.8\textwidth]{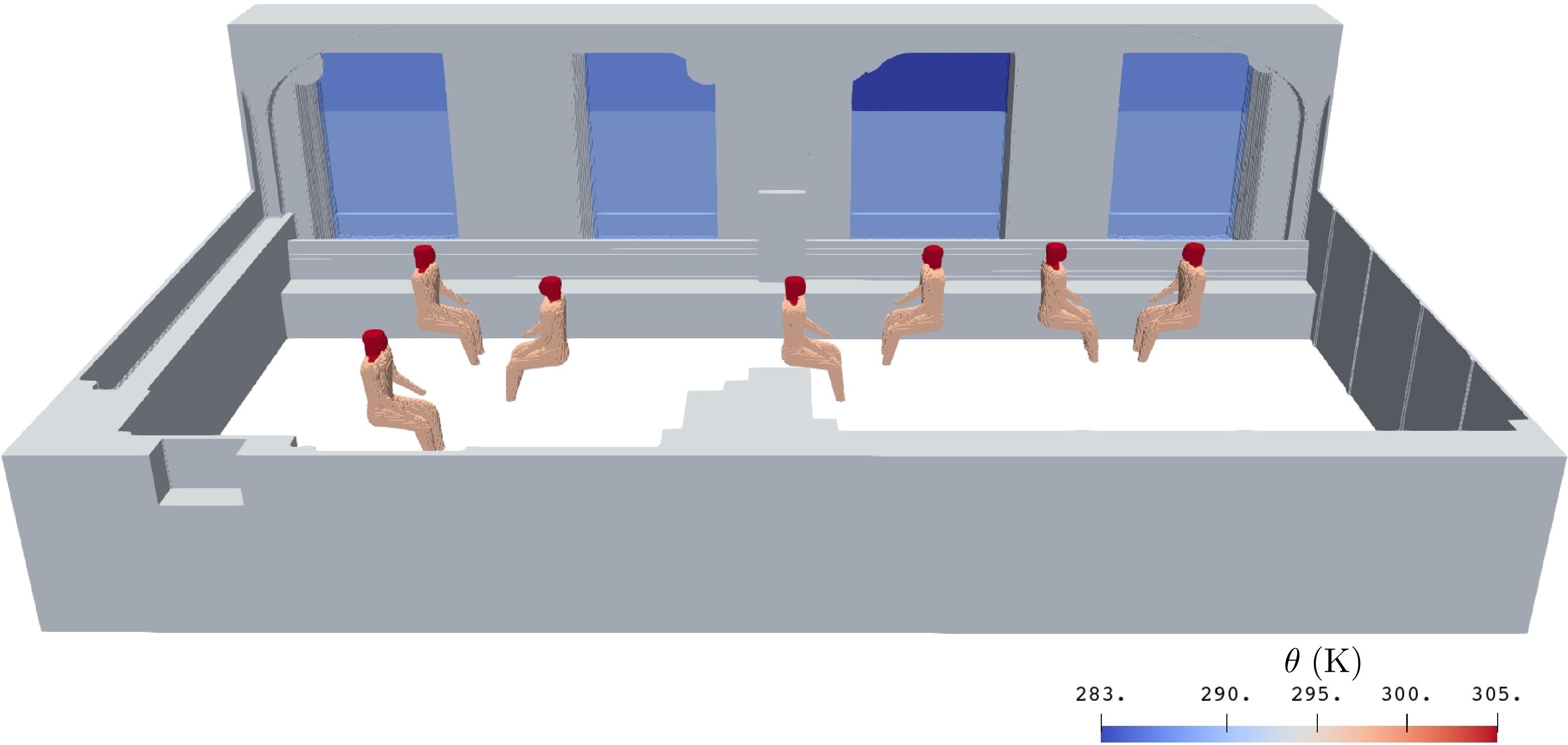}
    \caption{Cropped 3D visualization of the model specifying representative winter month surface temperatures for the room. The windows were assigned temperature distributions that were measured during the aerosol experiment. Human head and torso temperatures were estimated from infrared images. All other surfaces (some not shown) were set at room temperature $\unit[21]{^{\circ}\,C}$ ($\unit[294]{K}$).}
    \label{fig:les_thermal}
\end{figure*}

The winter time thermal complexity of the room, dictated by the heating radiators within the inlet ducts, cold window surfaces and the stationary humans occupying the room during the experiments, is captured by the model's thermal boundary conditions. The solid walls are considered to remain at room temperature, while the surface temperature of windows and clothed humans, measured on-site with a thermocouple and an infrared camera, are imposed as solid wall temperatures within the model. Figure~\ref{fig:les_thermal} depicts a clipped view of the 3D surface temperature input data model illustrating the thermal distribution of windows and humans. All other surfaces are set at room temperature. The room temperature is at $\unit[21]{^{\circ}C}$ ($\unit[294]{K}$) whereas the window temperatures were measured to be $\unit[16]{^{\circ}C}$ in the lower parts and $\unit[12-14]{^{\circ}C}$ in the upper parts. The mean surface temperature of human head was approximated to be at $\unit[32]{^{\circ}C}$ from infra-red images and the areas covered by clothing at $\unit[24]{^{\circ}C}$. Inlet ducts 1 and 3 held thermal radiators which heated the incoming air such that $\theta^{(1)}_{in} = 27~\mathrm{^{\circ}C}$ ($\unit[300]{K}$) and $\theta^{(3)}_{in} = 23~\mathrm{^{\circ}C}$ ($\unit[296]{K}$). The humans are included in the model to replicate the complex conditions found at real restaurants and to mimic the seating arrangement of a joint experiment which investigated the infectivity of airborne viruses (to be submitted). 

The boundary conditions for the scalar field $c = c(\xb, t)$ representing the concentration distribution of airborne aerosol particles are set such that only the flow field influences the distribution. Thus, at each $i^{\mathrm{th}}$ inlet $c|_{\xb \in A(i)} = 0$ is imposed whereas all solid walls and outlet boundary planes obtain their value from the nearest prognostic node (zero-gradient conditions). The concentration generation is implemented via $\Phi_{c}(x_i, y_j, z_k, t) = s_c \Delta x \Delta y \Delta z \Delta t$ with unit source $s_c = 1~\unit{s^{-1}\, m^{-3}}$ at every $i, j, k$ computational cell within a specified source volume $\Delta V_{c}$. In this study the volume for the aerosol concentration source is $\Delta V_c = (3\,\Delta x \times 3\,\Delta y \times 3\,\Delta z)$, which is centered where the mouth of the imaginary infected individual and the nebulizer outlet in the experiment are located.

\subsubsection{Numerical modeling of ventilation and filtration}\label{sec:numerical_vent_filt}

With the imposed momentum boundary conditions, the effective ventilation rate of the model restaurant room is set to $Q_{out} = Q_{in} = \unit[820]{m^3 \,h^{-1}}$. The ventilation rate is approximately matched with the average of the on-site measurements from the inlet and outlet ducts. In the real room the outlet flow rate is set to exceed the inlet flow rate by approximately 20\% in order to maintain the indoor pressure level slightly below the ambient pressure. The volume-flow balance becomes closed due to various leak flows from outside through e.g. non-ideally sealed windows and doors. The measured total volume flow rate at the restaurant outlets is approximately $\unit[900]{m^3 \,h^{-1}}$ and at the inlets approximately  $\unit[740]{m^3 \,h^{-1}}$. No leakages exist in the LES model, hence equal in- and out-flow rates must be specified as boundary conditions. 

The study employs two air purifiers \citep{UniqAir}, which are generically positioned within the room without prioritizing any particular seating position (as the location of the sick person is always unknown in public places). Each device is operated at $81\%$ of its maximum filtration capacity ($\unit[330]{m^3\,h^{-1}}$) which results in a combined filtration volume flow rate $Q_{flt} = \unit[540]{m^3\,h^{-1}}$. Thus, the air purifiers introduce a $65\%$ increase to the original ventilation capacity. Table~\ref{tab:ventilation} summarizes the relevant ventilation metrics of this study.

\begin{table}[ht]
\caption{Relevant ventilation metrics of the LES.}
\centering
\begin{tabular}{lcc}
 \toprule
  {}      & $ Q \; \unit{(m^3\,h^{-1})} $  &  $Q/V \; \unit{(h^{-1})} $ \\
  \midrule
   Base Ventilation:   &  820  & 4.8 \\
   Filtration:         &  540  & 3.2 \\
  Total Ventilation:   &  1360 & 8.0 \\
\bottomrule
\end{tabular}
\label{tab:ventilation}
\end{table}

\subsubsection{LES model runs}  \label{sec:les_runs}

The numerical simulations were initialized by running the indoor model for at least $\unit[45]{min}$ to sufficiently develop the thermal stratification and achieve a statistically steady-state flow conditions within the room. Simulations featuring air purifiers were restarted from base ventilation results which reduced the required spin-up time to $\unit[5]{min}$. The scalar concentration field was always zeroed before the source was activated again for the duration of each simulation. All simulations were run for $\unit[60]{min}$ which was determined to be a sufficient time scale for the concentration field to reach near asymptotic level within the room with base ventilation settings.

With third-order Runge-Kutta time integration approach, the maximum allowed $CFL$ number was set $CFL_{\mathrm{max}} = 1.2$, which is higher than the PALM default value 0.9, in order to reduce the already high computational cost of the simulation. Through testing it was determined that this increase in time step length did not affect the results. Here the time step length is dictated by the high speeds occurring either at the ventilation outlet duct or at the air purifier discharge jets where the maximum flow speeds reached around $\unit[4]{m\, s^{-1}}$ consistently. The average length of the resulting time step was $\overline{dt} = \unit[3.6 \times 10^{-3}]{s}$. The simulations were run with 864 CPU cores on the  Atos super cluster of CSC – IT Center for Science LTD featuring Intel Xeon Cascade Lake processors with 20 cores each running at $\unit[2.1]{GHz}$, see \url{https://docs.csc.fi/computing/system/#puhti}. Each simulation required approximately $\unit[6]{days}$ of clock time.

A set of shorter $\unit[10]{min}$ simulations were also performed to facilitate a more detailed analysis on the effect of turbulence mixing on the aerosol dispersion. These simulations were run with identical boundary conditions as the long runs, utilizing a fully developed solution as a restart.


\subsection{Validation methods and data sampling}  \label{sec:num_sampling}

For the validation study, the LES model with GEN configuration was utilized and the time-accurate concentration sampling was achieved by exploiting the \textit{user code} functionality of the PALM system \cite{Maronga2015}. Each of the 27 sensor locations (shown in Fig.~\ref{fig:sensor_layout}) were implemented as monitoring points which recorded the mean concentration within a sampling volume $\Delta V_s = (3 \Delta x \times 3 \Delta y \times 3 \Delta z)$ at every time step leading to a mean sampling frequency of $\unit[278]{Hz}$.

As the nebulizing unit's generation rate of suspended aerosols could not be experimentally verified, the experimental and LES model results were made comparable by normalizing the concentration values at each sensor location as
\begin{equation}
    c^{\mathrm{+}} = \frac{c}{  \langle \overline{c_{\textsc{r}}} \rangle_g  }
\end{equation}
where $\langle \overline{c_{\textsc{r}}} \rangle_g $ is the geometric mean of the reference case's (GEN) last $\unit[10]{min}$ (near asymptotic part) concentration averaged over the all available sensors. Thus, the value provides a global metric for the mean concentration level within the room, distributing the obtained $c^{\text{+}}$ asymptotes of the reference case around unity and the cases with added filtration (FLT) a degree below that level. 

The validation is quantified by three different metrics: Root-normalized means square error ($RNMSE$), fractional bias ($FB$) and correlation factor ($R$), which are defined for the time series $c^{\text{+}}$ obtained from the experiment (subscript \textsc{o} as in observation) and LES model (subscript \textsc{m}) as follows:
\begin{eqnarray}
  RNMSE 
    = & \sqrt{ \dfrac{ \overline{ \left( c^{\text{+}}_{\textsc{o}} - c^{\text{+}}_{\textsc{m}} \right)^2 }  }{ \overline{c^{\text{+}}}_{\textsc{o}} \;   \overline{c^{\text{+}}}_{\textsc{m}} } }   \label{eqn:RNMSE} \\[3pt]
   FB = & 
  \dfrac{ \overline{c^{\text{+}}}_{\textsc{o}}  - \overline{c^{\text{+}}}_{\textsc{m}}}{0.5 \left( \overline{c^{\text{+}}}_{\textsc{o}} + \overline{c^{\text{+}}}_{\textsc{m}} \right) }  \label{eqn:FB} \\[3pt]
   R = & \quad
     \dfrac{ \overline{ \left( c^{\text{+}}_{\textsc{o}} - \overline{c^{\text{+}}}_{\textsc{o}} \right) \left( c^{\text{+}}_{\textsc{m}} - \overline{c^{\text{+}}}_{\textsc{m}} \right) } }{ \sigma_{c^{\text{+}}_{\textsc{o}}}   \sigma_{c^{\text{+}}_{\textsc{m}}} }  . \label{eqn:R}
\end{eqnarray}

The evaluation metrics $RNMSE$ and $R$ require matching sampling frequencies from the two time series. Naturally, the higher sampling frequency ($\unit[278]{Hz}$) of the LES-predicted $c^{\text{+}}_{\textsc{m}}$ time series is filtered to meet the lower ($\unit[0.2]{Hz}$) frequency of the experiment. This is achieved by applying a local mean filter to $c^{\text{+}}_{\textsc{m}}$, such that the filter width matches the sampling time step of the aerosol sensor, and extracting values with matching time stamps for the metric evaluation. 

In addition to the performance metrics, the model evaluation also inspects the lag time  $\Delta t_0~=~(t_a - t_0)$ defined as the difference between the arrival time $t_a$, which is the instance when the concentration signal reaches 5\% of the local asymptote, and the activation time $t_0$ of the aerosol source. This travel time provides evidence on the mean flow within the room. 

The inspection of the performance metrics and the comparison of time lags at 27 spatially distributed sampling locations within the indoor domain, sets an indoor model validation standard that is both thorough and transparent and previously, to the authors' knowledge, not documented in indoor CFD studies.

\subsubsection*{Data output for 3-dimensional analysis}

For analyzing the spatial and temporal evolution of the system, the concentration and velocity field within the entire room interior of the LES model domain is saved at $\unit[2]{cm}$ resolution and at $\unit[0.05]{Hz}$ sampling rate (every 20 seconds). All the subsequent analysis are performed with post-processing scripts. For the shorter simulations, used in analyzing the effect of enhanced turbulent mixing (Section~\ref{sec:turbulence_mixing}), the sampling rate was increased to $\unit[1]{Hz}$.

\section{Results and discussion}   \label{sec:results}

\subsection{Model evaluation} \label{sec:model_evalution}

The qualitative assessment of the LES model validation is guided by the performance criteria employed, for example, by \citet{Moonen2013} in accordance with the more stringent air quality acceptance recommendations laid out in \citet{Chang2004}. At this point, adopting the acceptance criteria derived for \textit{air quality} models seems most appropriate due to the strong analogy between aerosol dispersion indoors and pollutant dispersion outdoors and the complexity of both physical systems. The acceptance ranges are $RNMSE < \sqrt{1.5} \, (0)$, $\left| FB \right| < 0.3 \, (0)$ and $R > 0.8 \, (1)$ indicating the ideal values in parenthesis. The metrics are chosen such that $RNMSE$ reveals the degree of both random scatter and systematic bias between the two datasets whereas $FB$ provides a specific indication of systematic bias. The correlation factor $R$, in turn, reflects the linear relationship between the two time series.  

\begin{figure*}[ht]
    \centering
    \includegraphics[width=0.8\textwidth,trim={0.0cm 0.0cm 0.0cm 0.0cm},clip]{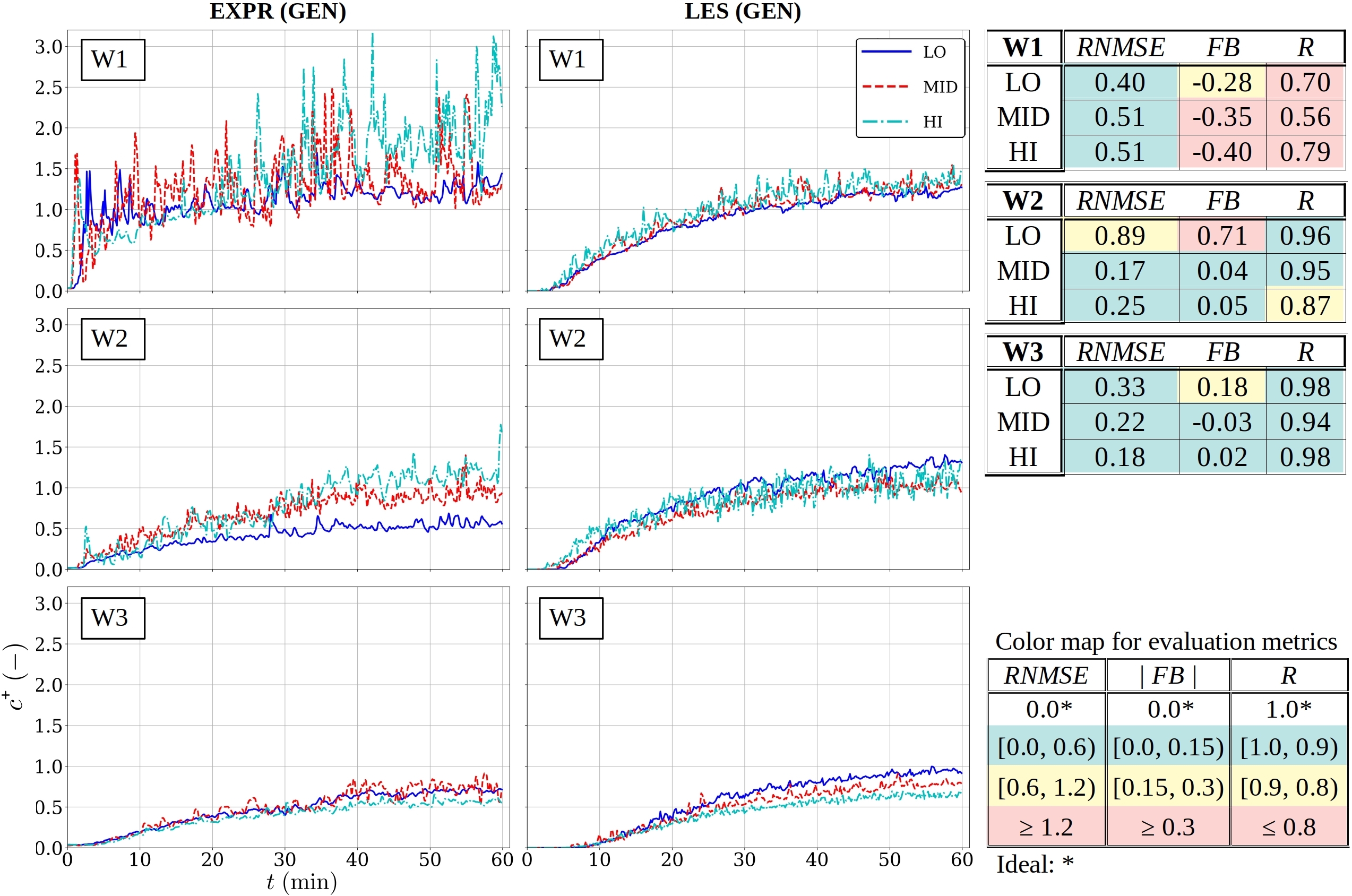}
    \caption{Comparison of measured (left) and modelled (middle) normalized concentration time series in the window-side mast row without the air purifiers (GEN) and tabulated evaluation metrics (right) with color coding. The color coding and the acceptance criteria are tabulated on the lower right corner.}
    \label{fig:expr_les_genW}
\end{figure*}

\begin{figure*}[ht]
    \centering
    \includegraphics[width=0.8\textwidth,trim={0.0cm 0.0cm 0.0cm 0.0cm},clip]{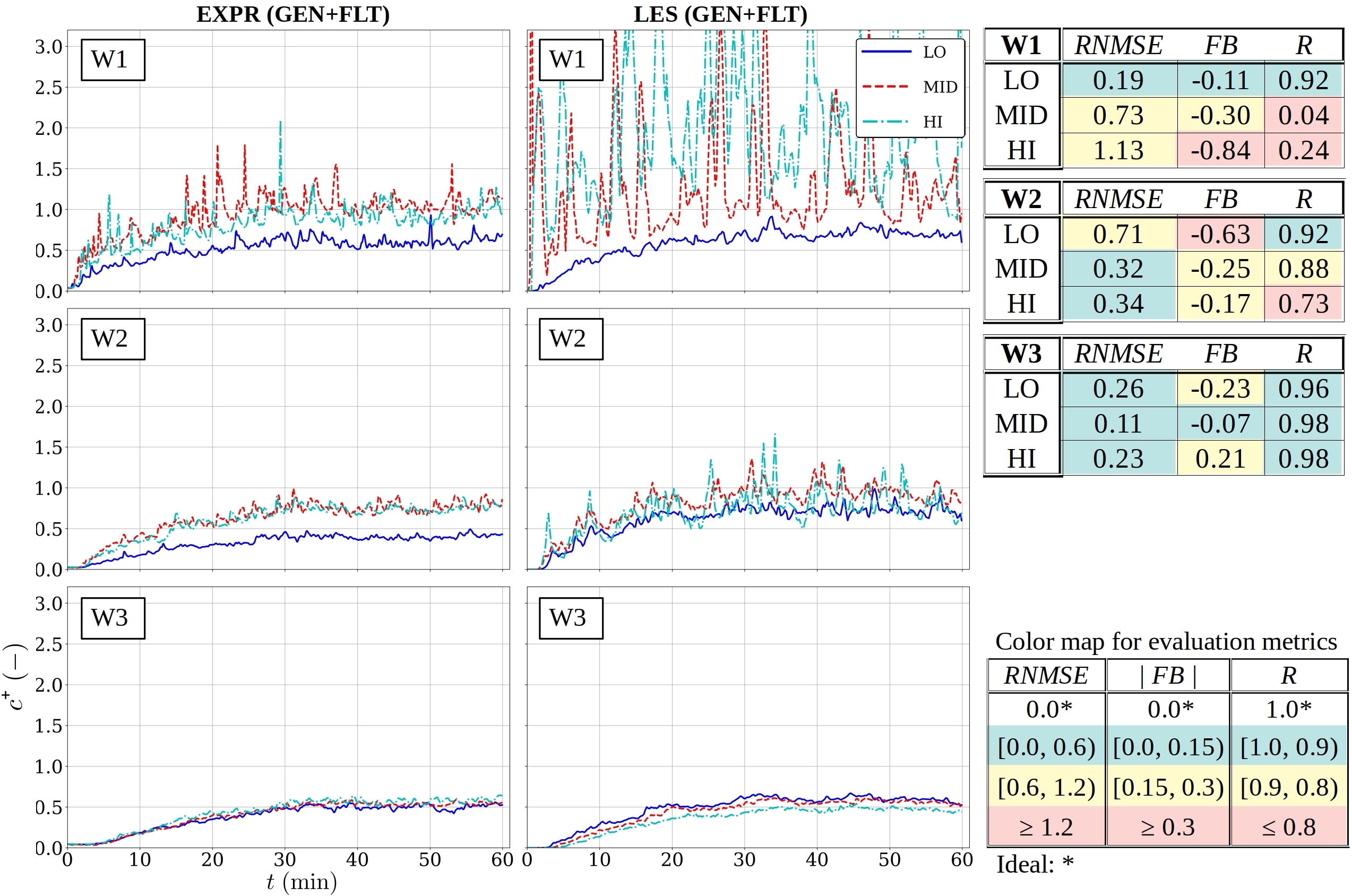}
    \caption{The same as in Fig.~\ref{fig:expr_les_genW} but with the air purifiers (GEN+FLT).}
    \label{fig:expr_les_fltW}
\end{figure*}

The model evaluation is presented with a series of plot compilations featuring a juxtaposition of measured and modelled $c^{\text{+}}$ time series and tabulated performance metrics for each pair. One such compilation contains all 9 time series obtained along a mast row, labelled either [W]indow, [C]enter and [D]oor as shown in Fig.~\ref{fig:sensor_layout}. Here, two compilation figures are presented for mast locations W featuring GEN (Fig.~\ref{fig:expr_les_genW}) and GEN+FLT (Fig.~\ref{fig:expr_les_fltW}) configurations. The other two mast locations (C and D) for both configurations are presented in Appendix~\ref{app:evaluation_results} Figs.~\ref{fig:expr_les_genC}-\ref{fig:expr_les_fltD}. For a complete summary of the model validation, all the performance metrics are gathered in Table~\ref{tab:valid_metrics}.

To aid the interpretation of the model performance, a traffic light color scheme is employed in highlighting the obtained metrics at each sensor location. The color map divides the acceptable range in two, labelling the half closest to the ideal value with green, the other half with yellow and the values falling outside the acceptable range with red. The color map for each evaluation metric is included on the bottom right-hand-side of every figure.     

A comparative inspection of the time series depicted in Figs.~\ref{fig:expr_les_genW} and \ref{fig:expr_les_fltW} (and Figs.~\ref{fig:expr_les_genC}-\ref{fig:expr_les_fltD} in the appendix) grants the following general observations: 
\begin{itemize}\setlength\itemsep{-2pt}
    \item The characteristic concentration growth curve is identifiable in all time series and it is well captured by the LES model results.
    \item The turbulent fluctuations (which are unique to every realization) are most observable closest to the concentration source (sensor masts W1 and C1) and they subdue considerably with distance from the source.
    \item With only few exceptions, the asymptotic levels are well reproduced. The largest deviations occur close to the source where the dispersion outcome is most sensitive to the nebulizer's plume behavior in the experiment which is influenced by positive buoyancy due to the heat generated by the nebulizer and negative buoyancy due to the cooling evaporation process that occurs when the plume is emitted. This thermal complexity of the aerosol concentration source is not replicated in the LES model where the conditions around the source are near neutral. 
\end{itemize}

The summary of the evaluation metrics in Table~\ref{tab:valid_metrics} lays bare that 100\% of the $RNMSE$ values register as acceptable (green or yellow) and 96\% of the values for the GEN and 81\% for the GEN+FLT configurations register at firmly acceptable level (green). This establishes the overall consistency of the time series, which is further substantiated by the high acceptance rate of the correlation factor $R$ (89\% for GEN and 69\% for GEN+FLT) which needs to take into account that the shortcomings (red values) are primarily due to the realization-specific turbulent fluctuations near the source (sensor masts W1, C1 and D1). From the perspective of these two metrics, the model validation is firmly supported. (D1\_LO time series is omitted from the GEN+FLT dataset because the signal was tainted by the air purifier's discharge jet.)

The fractional bias ($FB$) values, on the other hand, reveal that systematic bias  dominates the deviation between the experimental and model datasets as nearly half (48\%) of the $FB$ values for GEN and 42\% for the GEN+FLT configuration fall outside the acceptable range. The deviations in temporal mean values (see Eqn.~\ref{eqn:FB}) arise from two sources: the systematic offset in the asymptotic levels (e.g.\! W2-LO in Fig.~\ref{fig:expr_les_genW}) and the differences in the lag time $\Delta t_0$ measuring the arrival time of the concentration front (e.g.\! D3-LO in Fig.~\ref{fig:expr_les_genD}). 

\begin{figure*}[ht]
    \centering
    \includegraphics[width=0.8\textwidth,trim={1.2cm 6.2cm 6.6cm 0.5cm},clip]{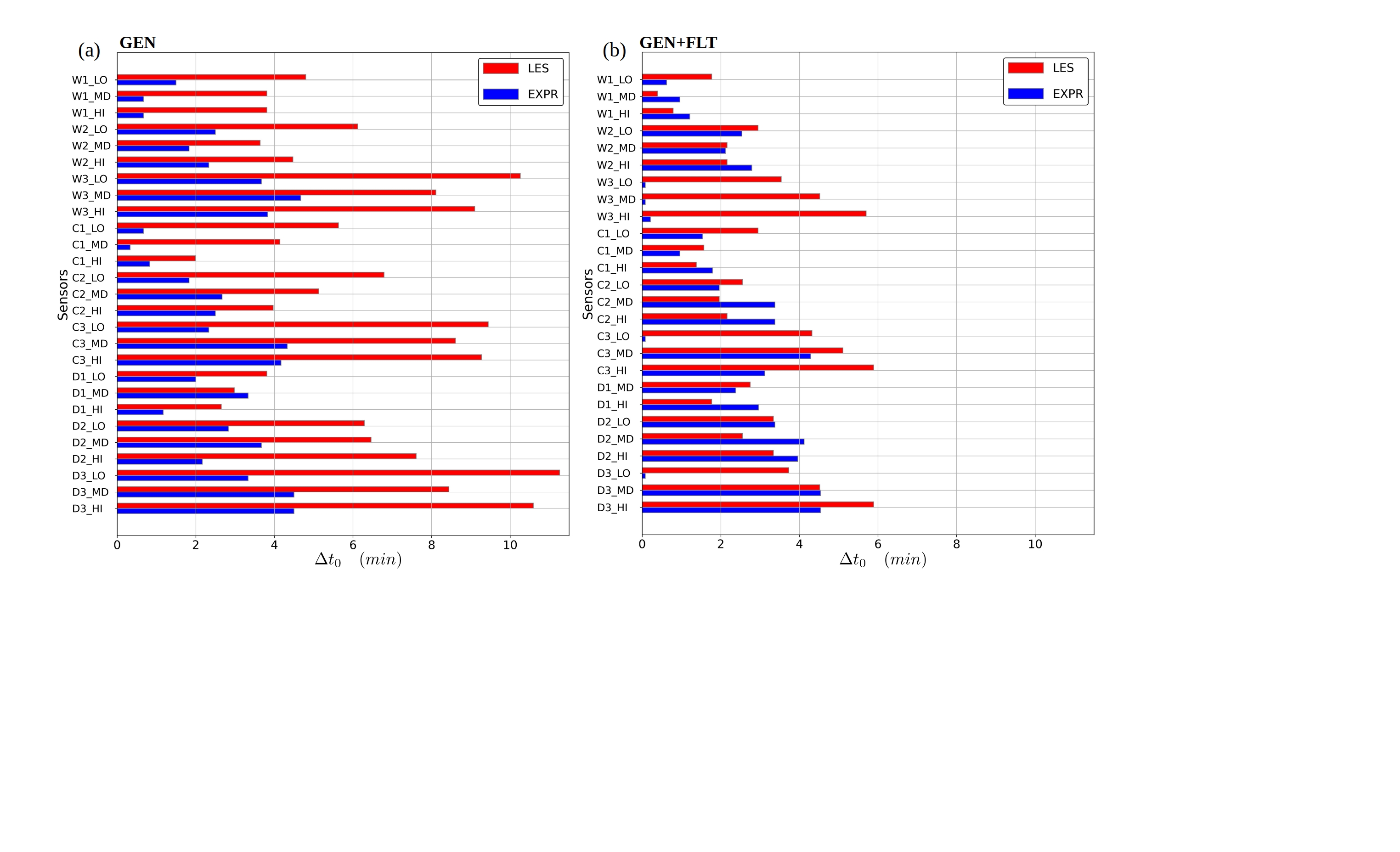}
    \caption{Bar plot comparison between the experiment and LES simulation of elapsed time $\Delta t_0$ to reach $0.05 \, c^{\text{+}}_{t \rightarrow \infty}$ at each measurement sensor. Subfigure (a) features for the generic ventilation configuration and (b) the configuration with air purifiers.}
    \label{fig:dt_init}
\end{figure*}

A comparison of the lag times shown in Fig.~\ref{fig:dt_init} reveal for the GEN ventilation configuration that the LES model systematically under predicts the advection of the initial concentration front. This arises from the intensified through-flow across the real room which is sustained by the leakage flows occurring at the windows causing an imbalance between the inlet and outlet flow rates. The ideally sealed LES model, whose boundary condition setup prioritizes the correct generation of indoor turbulence, is not equipped to capture such imperfections and therefore the observed systematic inadequacies in $FB$ values primarily reflect the complexity of a real-world indoor flow system than the inability of the LES model to capture the relevant indoor flow physics governing aerosol dispersion.  

However, the addition of the air purifier units (GEN+FLT configuration) to central locations within the room significantly energizes the flow, making the units a dominant source of momentum (particularly in the LES modelled flow system). From the flow's perspective, the air purifiers act as fans which add energy into the system. This added energy diminishes the LES flow solution's dependency on the inlet and outlet boundary conditions and their uncertainties, subsequently reducing the lag times $\Delta t_0$ in the LES results. The systematic reduction is made apparent by comparing the bar plots for GEN and GEN+FLT configurations in Fig.~\ref{fig:dt_init}. Due to the stronger through-flow in the experimental setup, the measured lag times did not exhibit similar sensitivity to the energized flow conditions.

\begin{table}[ht]
    \centering
    \caption{Collection of tables summarizing the validation metrics $RNMSE$, $FB$ and $R$ from all sensor locations for generic (GEN) and augmented filtration (GEN+FLT) ventilation configurations. The color map for the evaluation metrics is shown in the middle. Green and yellow indicate acceptance.}
    \label{tab:valid_metrics}
    \includegraphics[width=0.49\textwidth,trim={0 0 0 0.0cm},clip]{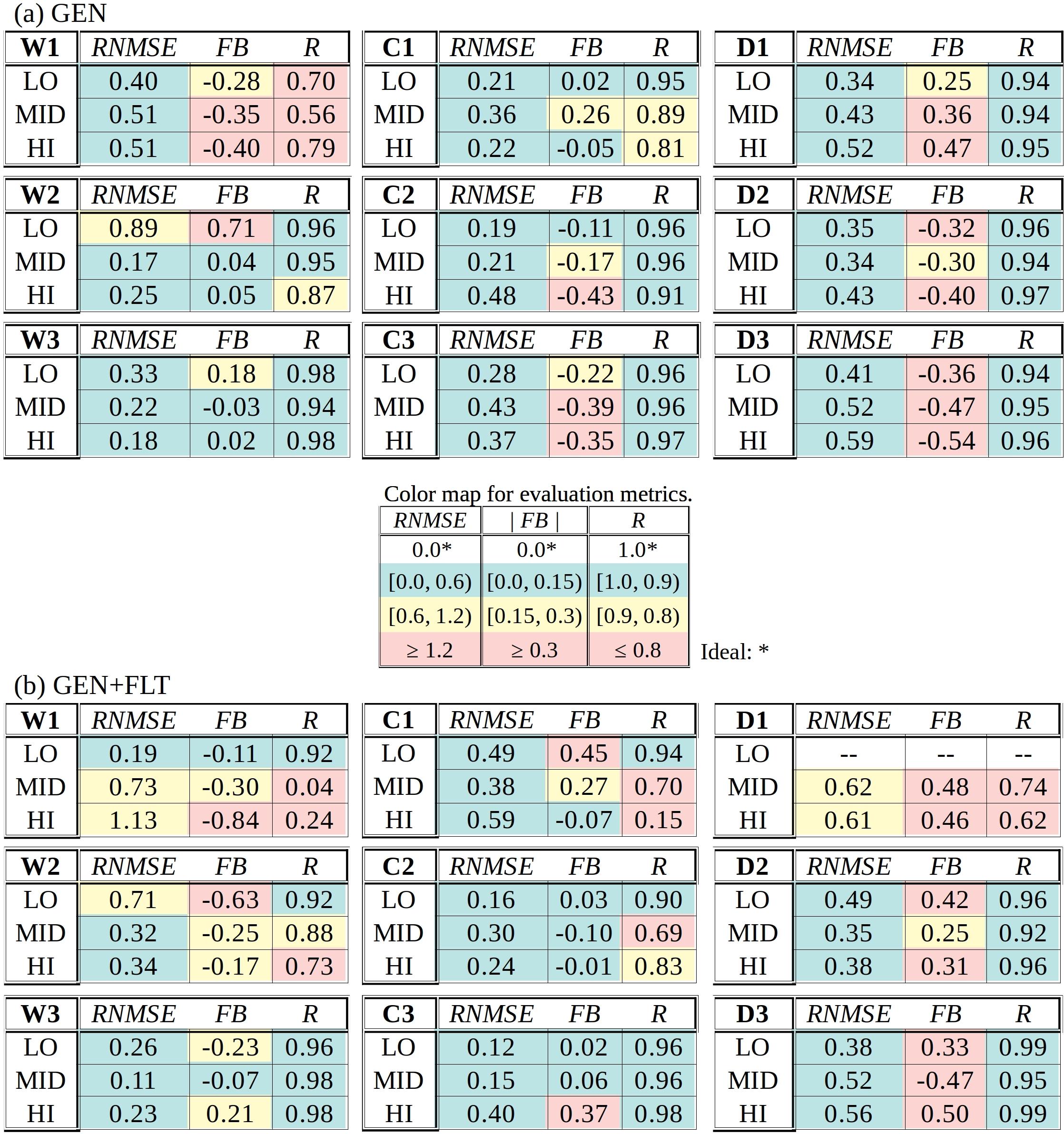}
\end{table}

It is essential to acknowledge that both experimental and LES model time series each represent a unique realization of a lone dispersion event with a continuous source. The turbulent nature of the flow system dictates that individual realizations will vary, particularly at the beginning of the dispersion event whereas the long-time asymptotic state is not expected to alter between realizations. Thus, the concentration signals are expected to embody a degree of randomness which decreases with time.  For most rigorous outcome, a statistically representative ensemble of realizations should have been analyzed, but with the currently available experimental and super-computing resources, this could not be achieved. Nonetheless, the level of detail and breadth of the validation enterprise is deemed high enough to provide sufficient evidence that the documented LES model is capable of resolving the relevant flow physics governing the spread of pathogen-carrying aerosols indoors.

\subsection{Indoor turbulence and dilution rate}  \label{sec:turbulence_mixing}

Here the test room's LES flow solutions are examined to gain insight on the physical phenomena which are relevant to this system specifically and indoor ventilation flows in general. It is acknowledged that implementation details of mechanical indoor ventilation arrangements vary between constructions, but the principal flow physics remain similar across different realizations. Principally, indoor flow systems are primarily taken up by free turbulence, which is maintained by mechanical and thermal momentum sources such as inlet jets and thermal plumes from heating or cooling elements.  The ventilation inlets are ideally designed to operate as jets which penetrate into the surrounding air, generating shear layers which diffuse via turbulent mechanisms, mixing freshly supplied air
uniformly throughout the living space. The outlets, on the other hand, are ideally placed such that the incoming air has the opportunity to mix into the ambient volume and sweep across the intended portion of the space before reaching the outlets and exiting the system. A ventilation short-circuiting, a process where freshly supplied air is exhausted before it has had the opportunity to mix sufficiently with the volume of air, is considered a major flaw in the system design. 

As ventilation flows are notoriously difficult to model and optimize, it is reasonable to expect that imperfections are ubiquitous in practise. The sensitivity of ventilation flows to modeling errors demands that the chosen numerical scheme is sufficiently accurate and robust that it is able to capture and uncover any undesired features in the system. This further highlights the importance of relying on high-resolution LES modeling which is capable of describing even the weaker turbulent structures without risking contamination due to modeling error. In effort to meet the stringent numerical demands, the LES model documented herein resolves at least 98\% of the turbulent kinetic energy within the space dominated by free turbulence. (The inlet and air purifier jets are partly under-resolved and thus the jets contain the greatest bulk of the subgrid-scale turbulent kinetic energy.) Thereby, it is evident that vast majority of the turbulent motion within the model is practically fully resolved, as in direct numerical simulation (DNS) approach. This is computationally feasible thanks to the exceptional parallel and cache efficiency of the structured PALM LES solver.

\begin{figure}[ht]
    \centering
    \includegraphics[width=0.3\textwidth]{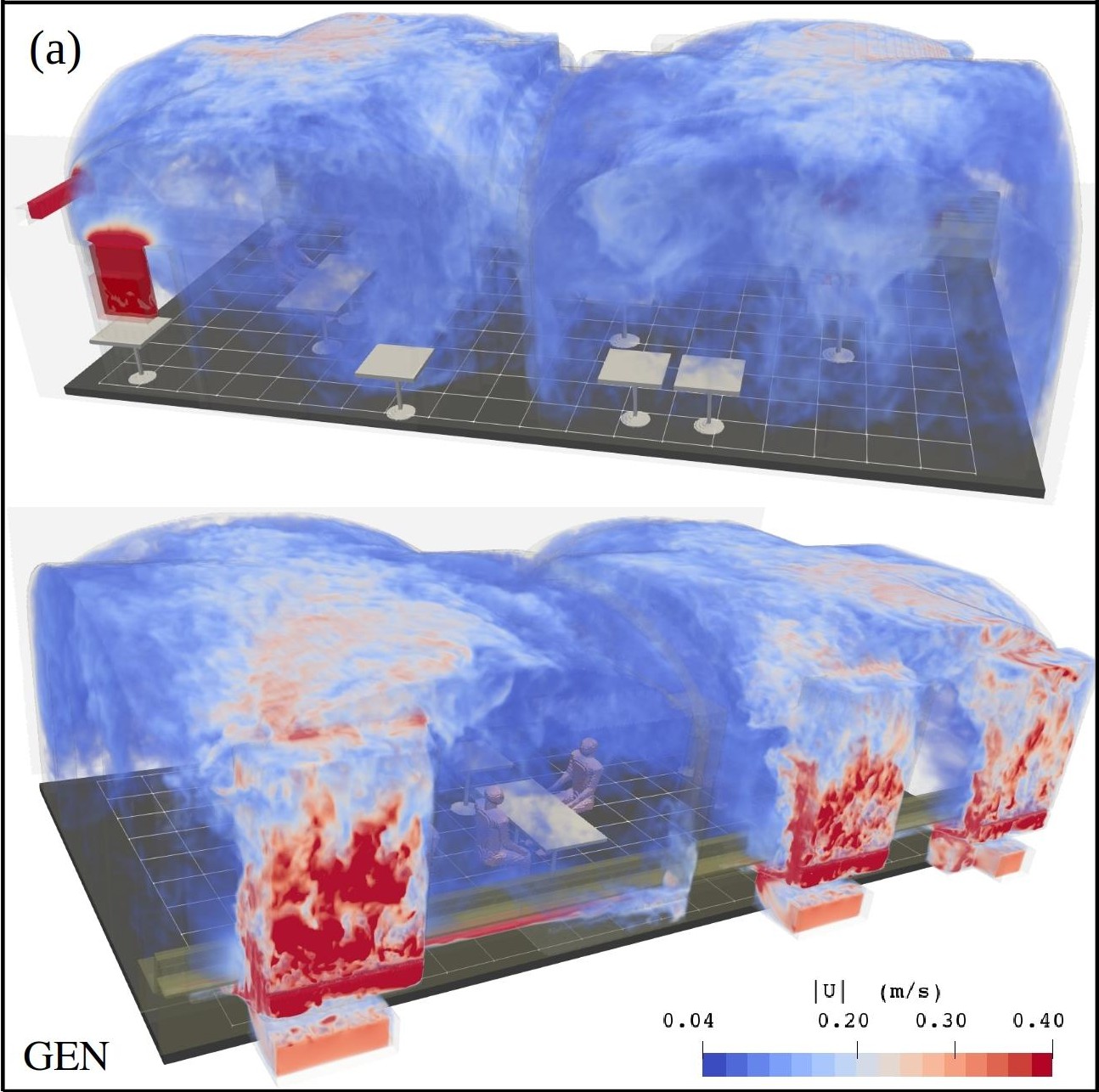}\\[-1pt]
    \includegraphics[width=0.3\textwidth]{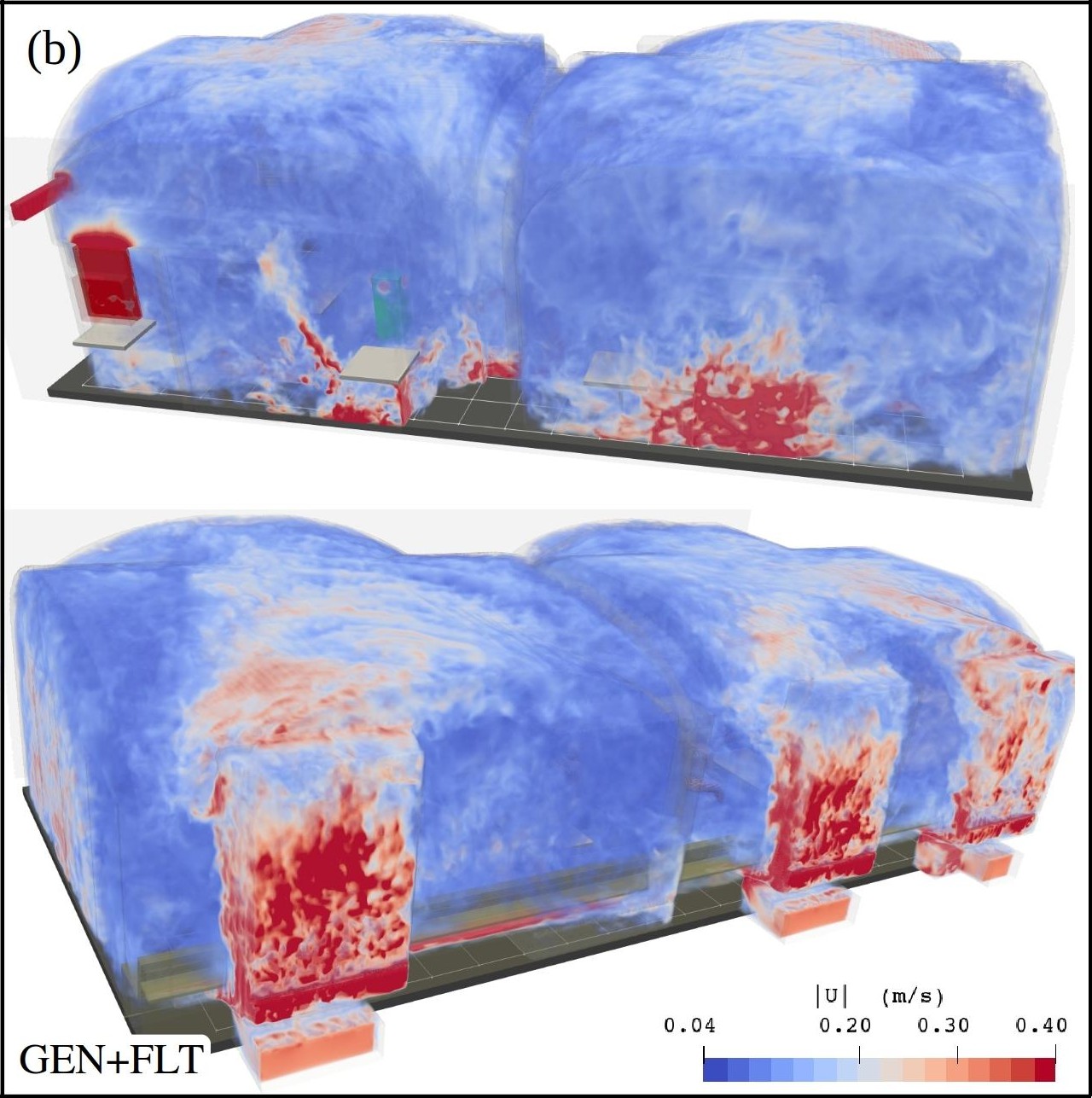}
    \caption{Volume rendering of instantaneous $\left| \bf{u} \right|$ field within the domain using two different view angles. Subfigure (a) depicts the GEN configuration and (b) the GEN+FLT configuration with added air purifiers. Note that regions where $\left| \bf{u} \right| < \unit[0.04]{m\, s^{-1}}$ are shown transparent revealing how the air purifiers intensify the flow system. The discharge jets of the air purifiers are visible in (b) as they become incident with the side walls.}
    \label{fig:umag}
\end{figure}

\begin{figure}[ht]
    \centering
    \includegraphics[width=0.3\textwidth]{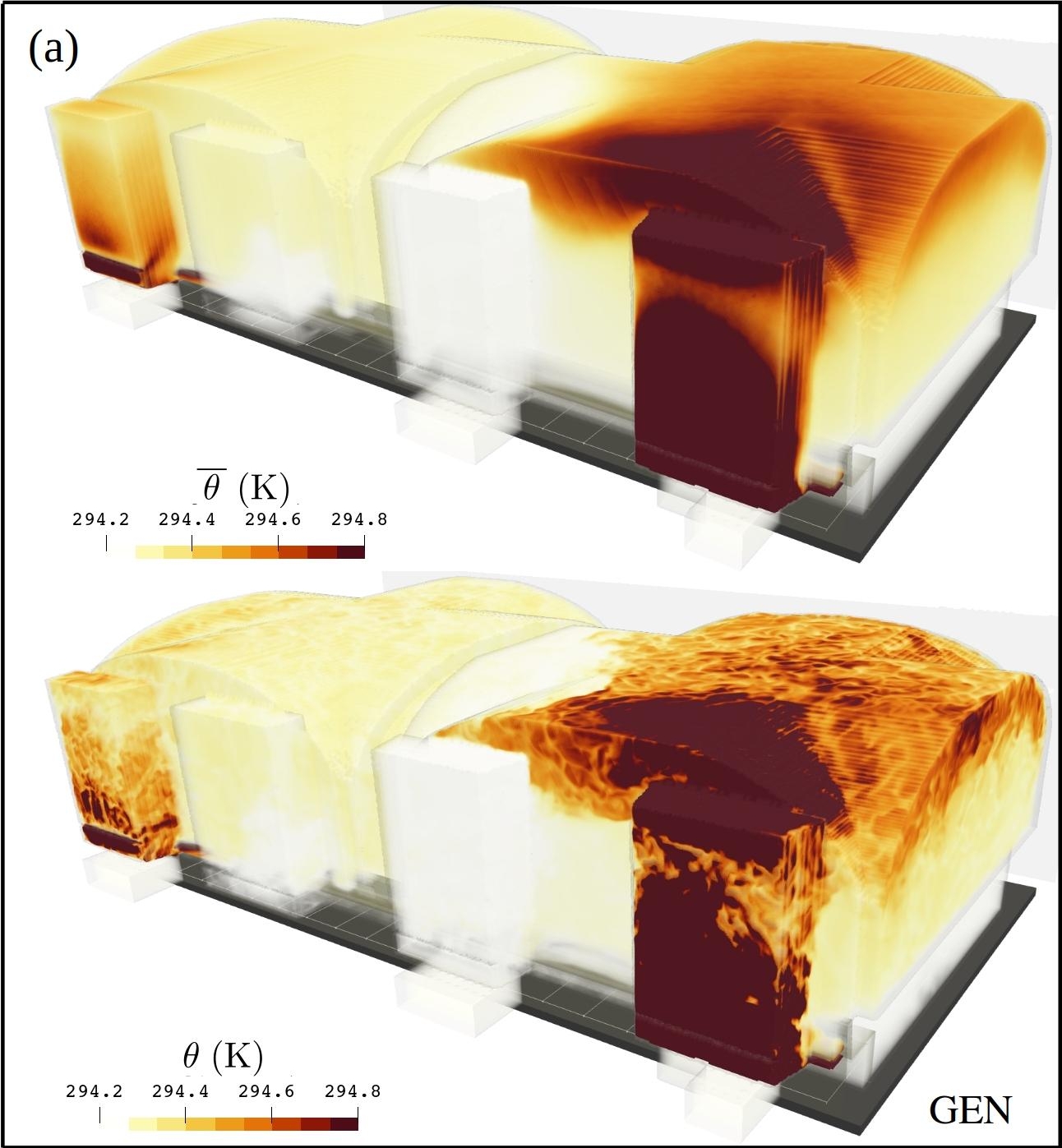}\\[-1pt]
    \includegraphics[width=0.3\textwidth]{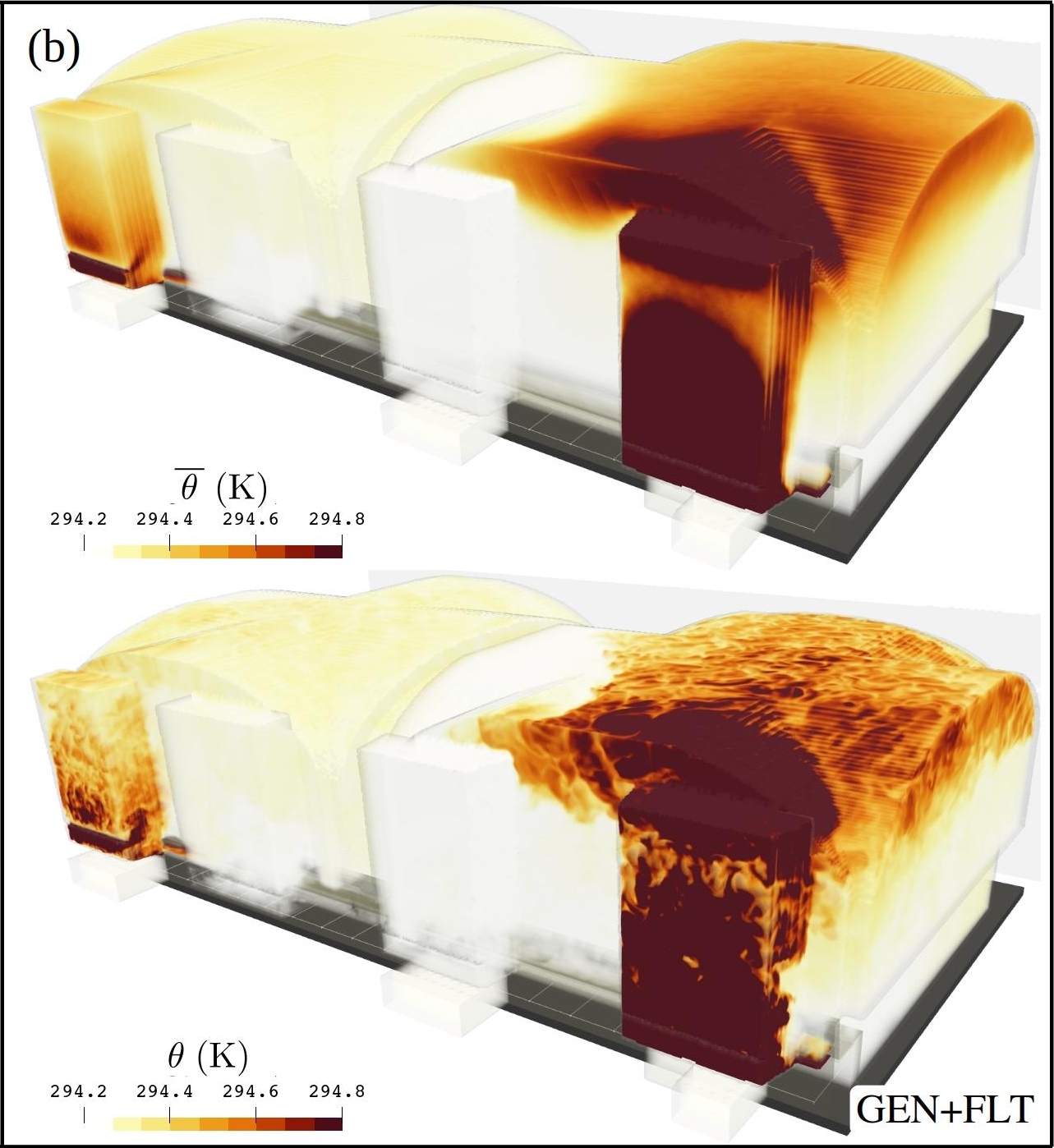}
    \caption{Visualization of mean and instantaneous temperature distributions for (a) GEN and (b) GEN+FLT configurations. The volume averaged mean temperatures are $\langle \overline{\theta} \rangle = \unit[294.35]{K}$ and $\unit[294.46]{K}$ for GEN and GEN+FLT configurations respectively. The most notable temperature gradients are introduced by the radiators within 1. and 3. inlet ducts and the cool windows.}
    \label{fig:temperature}
\end{figure}

The LES results reveal that the flow system of the test room does exhibit design flaws as undesired stratification is observed in both ventilation flow and temperature fields. An instantaneous state of the LES solved velocity magnitude field $\vert \bf{u} \vert$ is visualized in Fig.~\ref{fig:umag}(a) for GEN and (b) for GEN+FLT ventilation configurations. This three-dimensional flow speed distribution lays bare how, under the generic conditions, the flow energy is not evenly distributed throughout the room. Instead, the flow is amplified near the ceiling and systematically weaker ($\unit[<0.04]{m \, s^{-1}}$) near the floor. This imbalance is mutually supported by the room's thermal stratification which, while not strong, is apparent from the temperature distributions, both time averaged and instantaneous, shown in Fig.~\ref{fig:temperature} for (a) GEN and (b) GEN+FLT configurations. A thermal stratification of $\unit[2]{^{\circ} C}$ was verified with in situ measurements, but only after the room was allowed to lay dormant $\unit[>1]{h}$. Any human activity (movement) within the room disturbed the stratification reducing it below one degree. 

\begin{figure}[ht]
    \centering
    \includegraphics[width=0.35\textwidth]{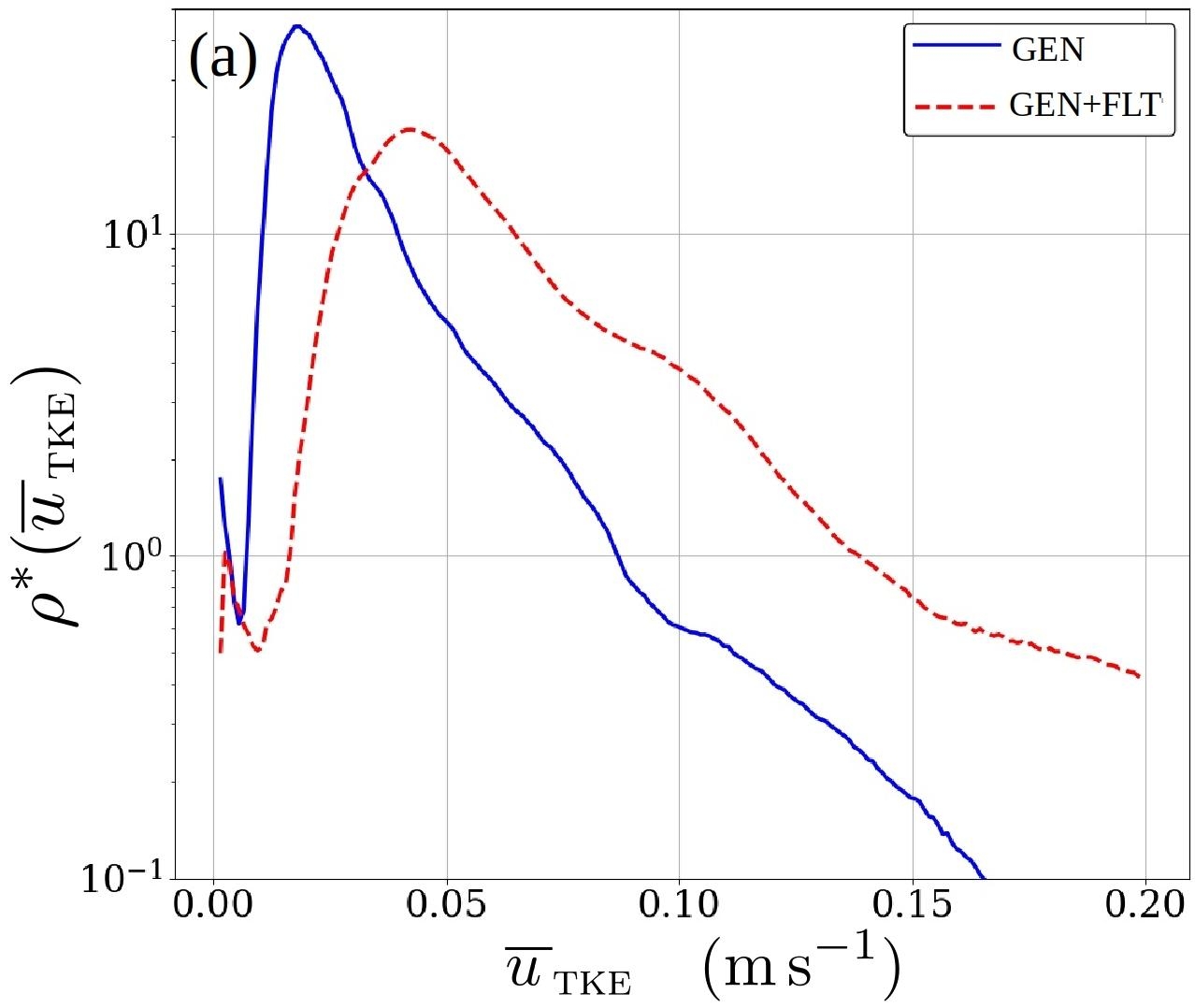}\\[2pt]
    \includegraphics[width=0.35\textwidth]{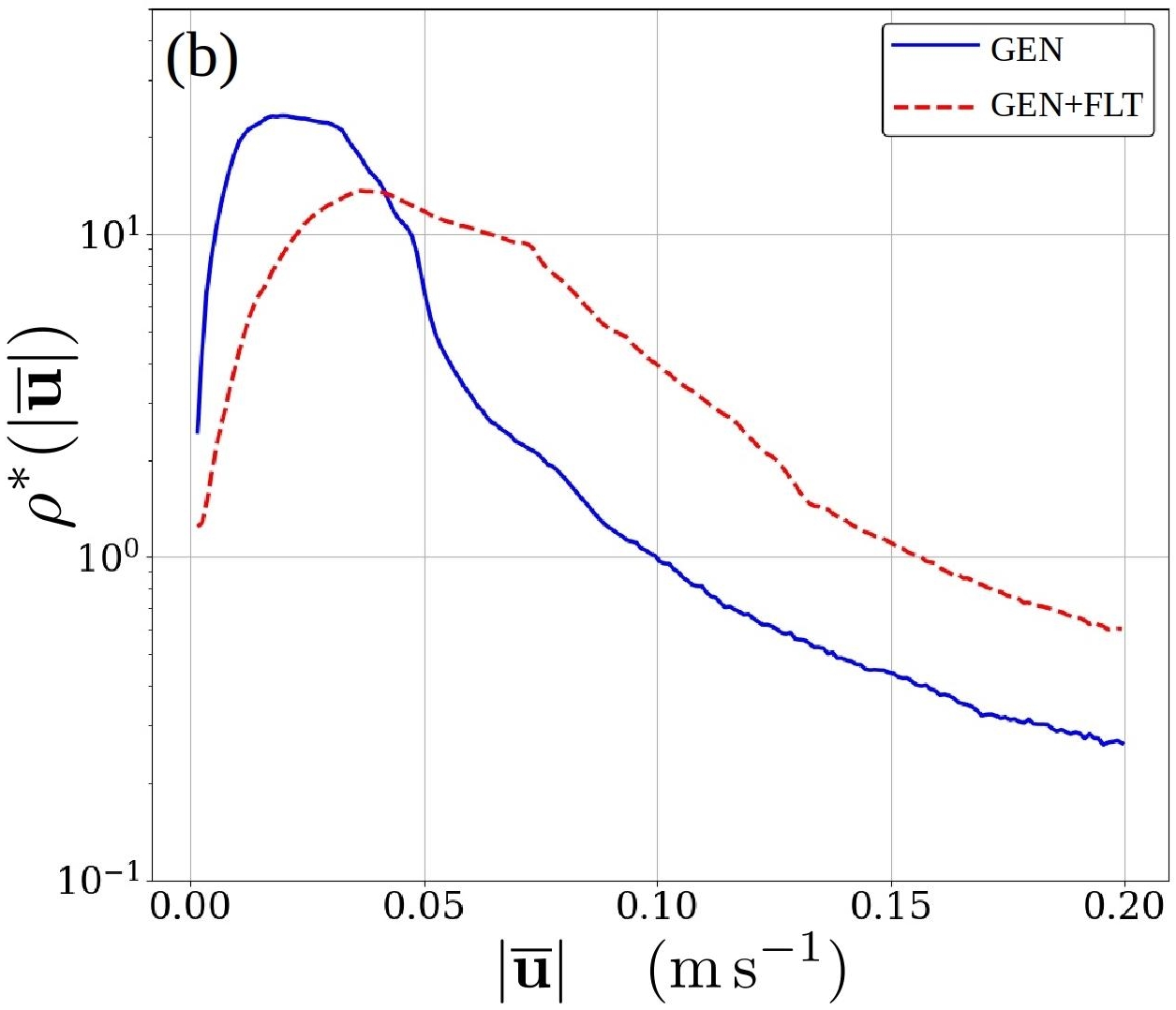}
    \caption{Probability density distributions $\rho^{*}$ of (a) mean turbulent kinetic energy speed $\overline{u\,}_{\textsc{tke}}(\xb)$ and (b) mean flow speed $\vert \overline{\bf{ u }}\vert (\xb)$ for all $\xb$ within the indoor LES domain. The shown speed range is focused on the characteristic indoor ventilation flow range outside the immediate vicinity of inlet and outlet ducts.}
    \label{fig:u-distributions}
\end{figure}

The introduction of air purifiers, which from the indoor air's perspective act as fans, significantly energizes the flow within the room, particularly in the stagnant lower part. This is visually apparent in Fig.~\ref{fig:umag}(b) and further quantified in Fig.~\ref{fig:u-distributions} depicting the probability distributions of mean turbulent kinetic energy speed $\overline{u\,}_{\textsc{tke}} = \sqrt{ (u'^2 + v'^2 + w'^2)/2 }$ and the mean velocity magnitude $\vert \overline{\bf{u}} \vert$ within the indoor domain. The distributions reveal how weak turbulence and mean flow dominates the system and how the addition of two air purifiers (effectively acting as small fans) shifts the distribution toward higher turbulent energies and mean flow speeds. This has consequences on the dilution properties of the flow system.

\begin{figure}[ht]
    \centering
    \includegraphics[width=0.35\textwidth]{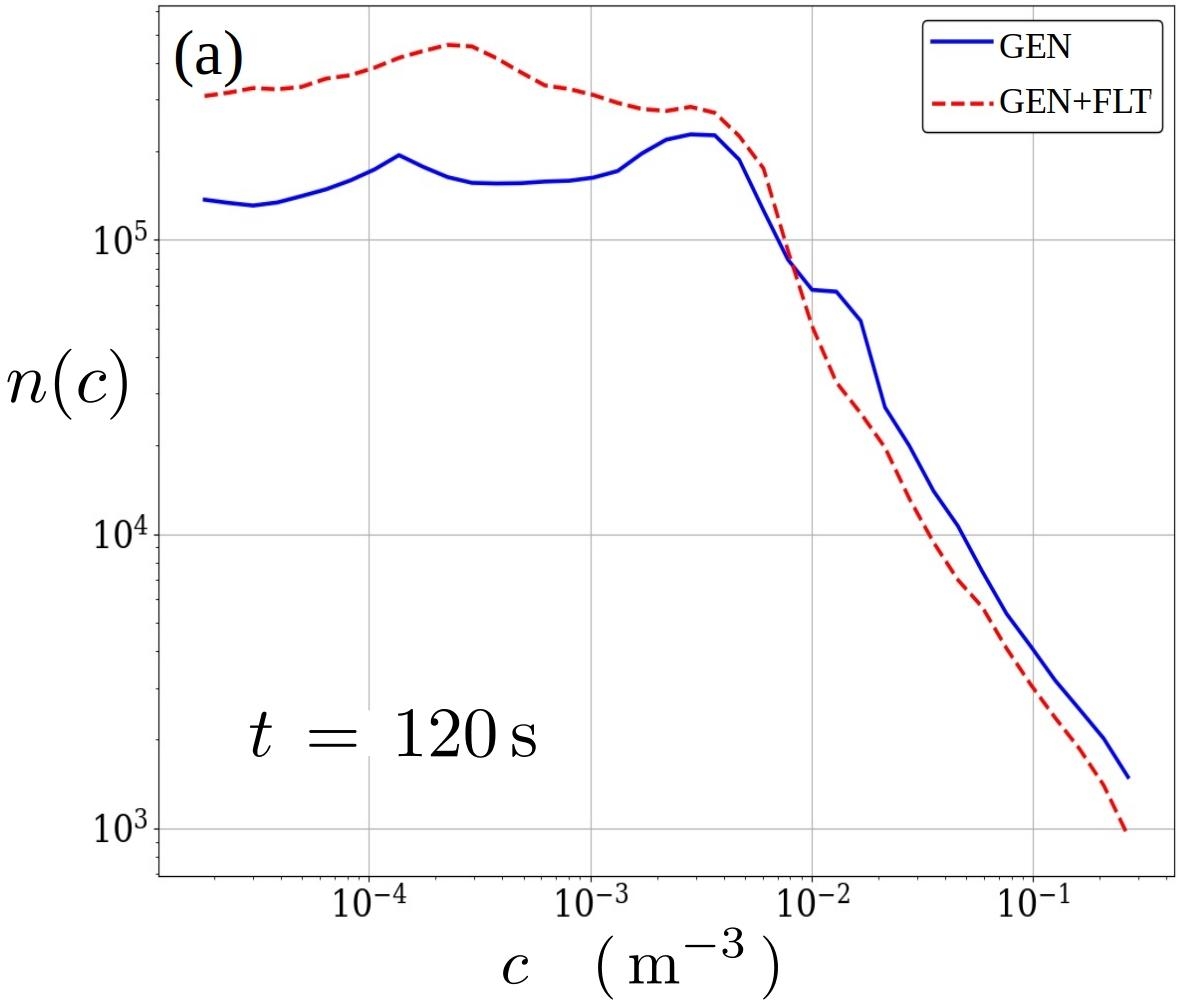}\\[2pt]
    \includegraphics[width=0.35\textwidth]{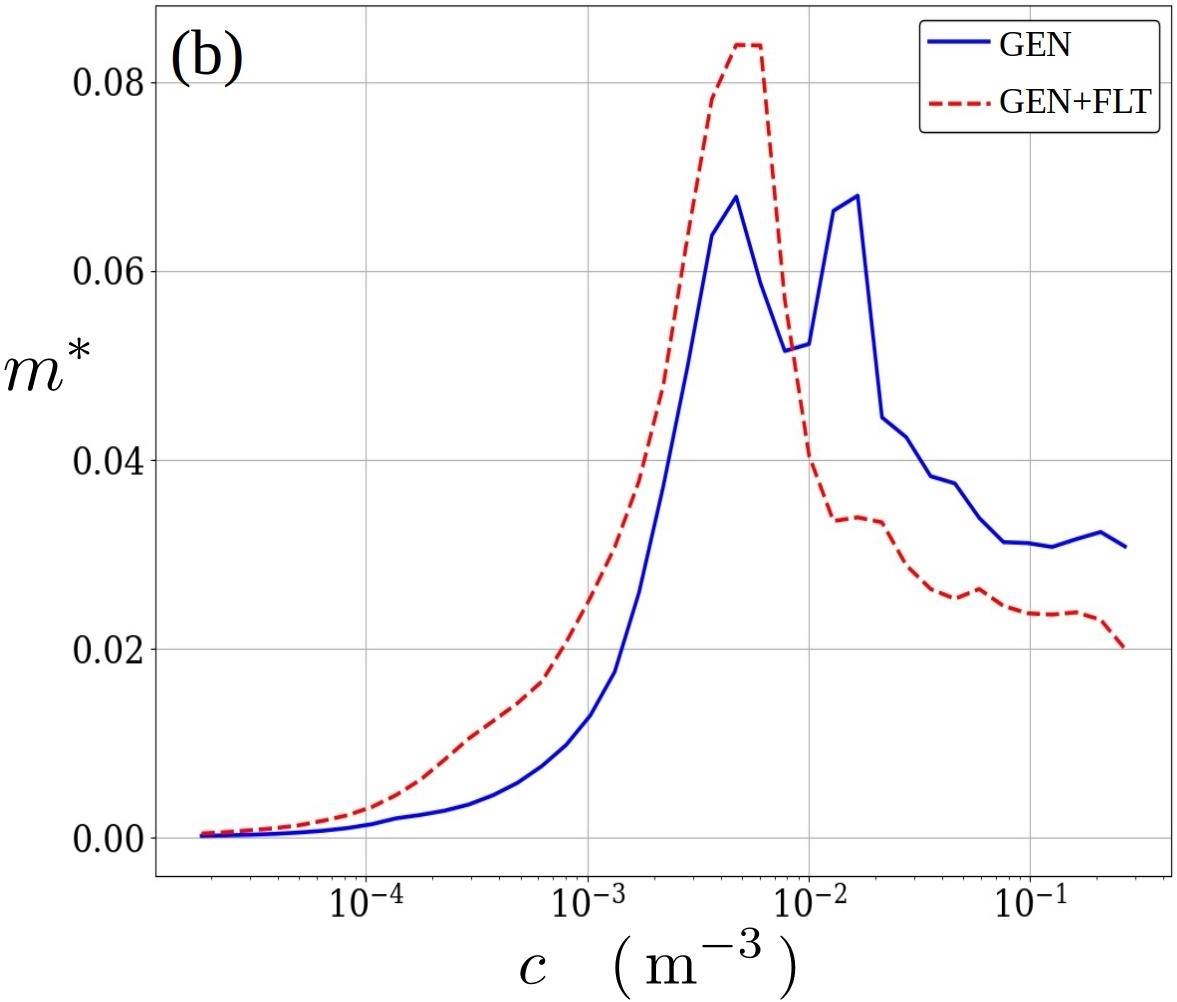}
    \caption{Distributions for (a) the number of occurrence $n$ and (b) the normalized scalar content $m^*$ for relevant concentrations at time instance $t = \unit[120]{s}$. At this instance, the differences in distributions are not due to filtration but solely due to air purifiers' influence on indoor turbulence level. The improved dilution rate arising from higher turbulence level in GEN+FLT reduces high concentration levels and shifts the scalar content distribution toward low concentrations faster.}
    \label{fig:cm-distributions}
\end{figure}

In the context of rapidly released puff of respiratory aerosols (due to a cough or a sneeze) the dilution rate of the aerosol concentration has been shown to scale with the turbulence level of the indoor flow system \citep{Vuorinen2020}. This diluting property of turbulence has been overlooked in the context of airborne transmission risk analysis. Thus, its role in a continuous emission event, taking place over a longer period of time, akin to a restaurant visit, is clarified herein by examining Fig.~\ref{fig:cm-distributions} illustrating a comparison between GEN and GEN+FLT configurations of how (a) the frequency (number of occurrence $n(c_i)$) of every $i^{\text{th}}$ concentration bin $c_i$ and (b) the normalized scalar content (i.e.\ mass fraction) $m_i^* =  n(c_i) \, c_i \, \Delta V / \sum_j (n(c_j) \, c_j \, \Delta V) $ is distributed within the system at $\unit[120]{s}$ after the onset of the aerosol source. The chosen time period is sufficiently long for the concentration field to disperse into a wide distribution of concentration values, but too short for the concentration field to reach the air purifiers. Thus, the resulting differences are solely due to the different flow conditions within the room.

The frequency distribution in Fig.~\ref{fig:cm-distributions}(a) clearly reveals how the elevated turbulence level has increased the occurrence of low concentrations, while subsequently reducing the higher concentrations. The most notable differences in the slope of the curves occur around range $\unit[7\times10^{-3} \ldots 3\times 10^{-2}]{m^{-3}}$ where the GEN+FLT configuration with air purifiers exhibits a faster rate of decline toward higher concentrations and does not feature the slight accumulation (bulge) seen in GEN curve around $\unit[1 \ldots 3\times 10^{-2}]{m^{-3}}$ range. These differences are readily visualized in the comparison of mass fraction distributions, which also confirms the effect of faster dilution rate by exhibiting an observable shift toward lower concentration values and the dissolution of the second peak found around $\unit[1 \ldots 3\times 10^{-2}]{m^{-3}}$ range in GEN results. Thus, higher turbulence level effectively dissolves peaks in the concentration field. It also distributes the concentration field around the space faster, which allows the outlets and added filtration to act on it faster. This leads to a clear reduction in overall concentration levels, bringing only positive consequences in the context of pathogen transmission probabilities which will be subsequently analyzed in Section~\ref{sec:analysis}.

\subsection{Infection probability} \label{sec:analysis}

Recently, infection probability has been studied in several works e.g. by \citet{Buonanno2020b, Buonanno2020a}. They applied the time-dependent form, introduced by \citet{Gammaitoni1997, Gammaitoni1997b}, of the original Wells-Riley model \citep{Riley1978}, but without any spatial variability effects due to the perfect mixing assumption behind this model. This limitation is remedied by combining CFD-predicted space-dependent concentration data with the Wells-Riley probability model in several studies \cite{Qian2009, Gupta2012, Yan2017, You2019, Guo2021, Foster2021}. However, none of these studies include mathematical or statistical justification for doing so despite the fact that the Wells-Riley probability model is based on the assumption of spatially constant concentration. This is amended by the subsequent novel treatment where we show how the Wells-Riley probability model can be extended for realistic, spatially varying CFD-predicted concentration fields. We will also show that, compared to the detailed LES-based prediction, the analytical model is likely to underestimate the infection probability by a significant margin. 

The analytical model for spatial average of the time dependent quanta concentration $\langle c(t)\rangle$ is a simple linear source-sink model \citep{Gammaitoni1997,Gammaitoni1997b}  
\begin{equation}\label{eq:analytical_eq_C}
  V\frac{\mathrm{d}\langle c\rangle}{\mathrm{d}t} = G_q - Q_{\mathrm{eff}}\langle c\rangle. 
\end{equation}
Here $V$ is the volume of the indoor space, $G_q$ is the quanta generation rate (quanta/s) and $Q_{\mathrm{eff}}$ is the effective air exchange rate including the volume-flow rate through the air purifiers if present (here we assume the purification efficiency is 100\%).  
Eqn.~\eqref{eq:analytical_eq_C} is based on the assumption of perfect and immediate mixing of quanta concentration within the space, i.e quanta concentration is assumed spatially constant. In our view, this assumption imposes a detrimental limitation on the model for three reasons. First, under real indoor ventilation conditions the mixing is primarily carried out by free turbulence which occurs gradually and leads to highly variable concentration fields. Second, under spatially varying concentration fields, it is highly improbable that the ventilation outlets remove concentrations at mean value, as stated by the sink term in Eqn.~\eqref{eq:analytical_eq_C}. Third, the assumption of constant concentration removes all peak concentrations within the space, which may lead to strong underestimations of the local infection probability.

\begin{figure}[ht]
    \centering
    \includegraphics[width=0.42\textwidth]{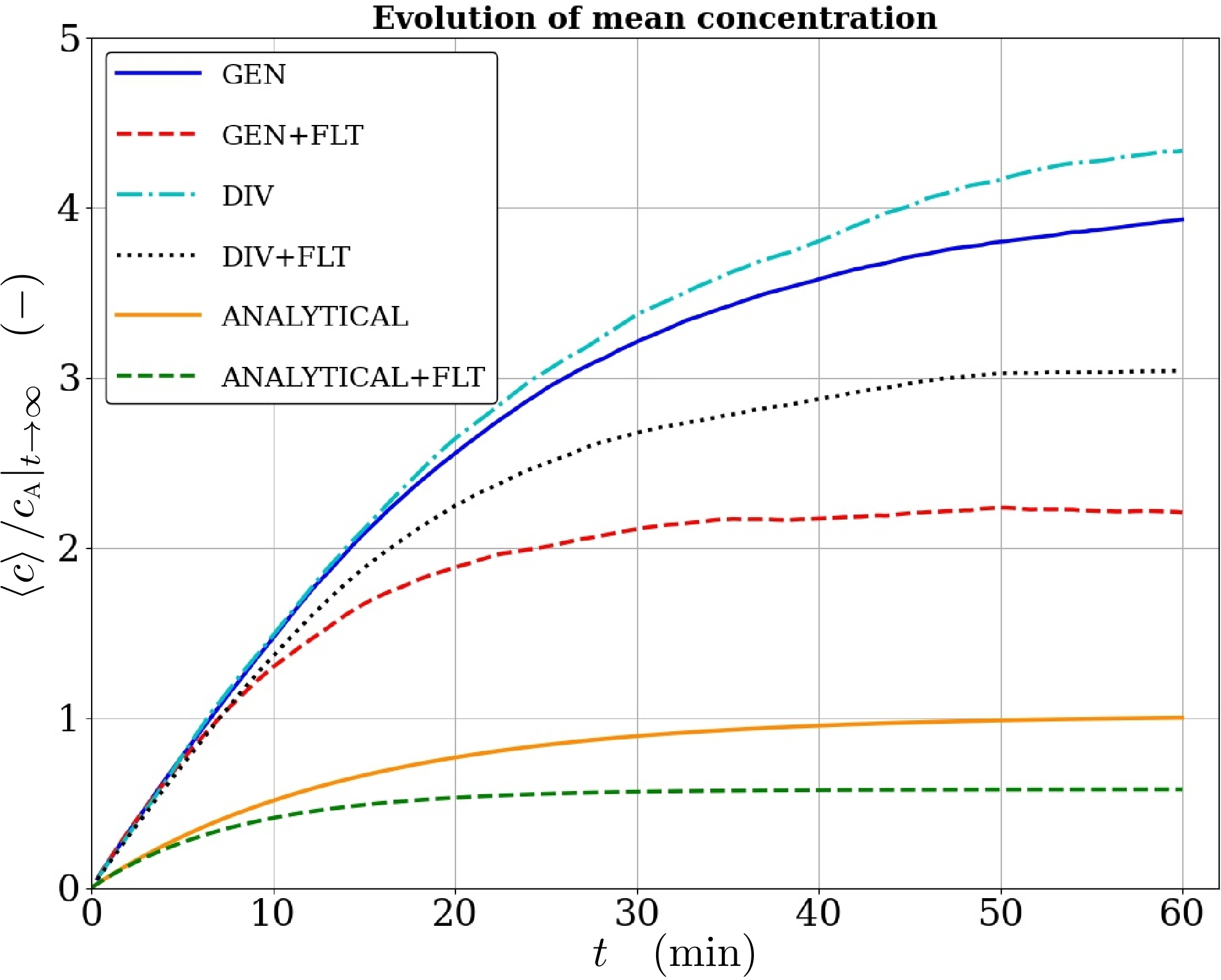}
    \caption{Temporal evolution of normalized spatially averaged concentrations according to LES without (GEN) and with (GEN+FLT) air purifiers, and with space dividers (DIV) and with both space dividers and air purifiers (DIV+FLT) compared with the analytical solution~\eqref{eq:analytical_sol_C} without and with air purifiers. All results are normalized by the the asymptote of the analytical solution $c_A\vert_{t \rightarrow \infty} = G_q/Q_{\mathrm{eff}}$ without air purifiers.}
    \label{fig:mean_evolution}
\end{figure}

Underestimation of the mean concentration predicted by the Wells-Riley theory in this case is clearly seen in Fig.~\ref{fig:mean_evolution} which shows the differences between the temporal evolution of scaled mean concentration predicted by the LES model and the scaled analytical concentration (utilized by the Wells-Riley model), i.e. the solution of Eqn.~\eqref{eq:analytical_eq_C} 
\begin{equation}\label{eq:analytical_sol_C}
  \langle c(t)\rangle = \frac{G_q}{Q_{\mathrm{eff}}}
         \left(1 - \exp{\left(-\frac{Q_{\mathrm{eff}}}{V}t \right)} \right) 
\end{equation}

Concentration results without (GEN) and with (GEN+FLT) the air purifiers are shown in Fig.~\ref{fig:mean_evolution}. The LES results include also the space divider cases DIV and DIV+FLT. Unlike the Wells-Riley model, the LES model resolves the complicated indoor turbulence which in turn dictates the spatial and temporal evolution of aerosol concentrations. All results in Fig.~\ref{fig:mean_evolution} are normalized by the long-time asymptote of the analytical solution (without air purifiers) $c_A \vert_{t \rightarrow \infty} = G_q/Q_{\text{eff}}$.   

The time dependent probability of infection $\langle P(t)\rangle$ is derived by \citet{Gammaitoni1997b}, see Appendix~\ref{app:P_extension}. Here, we present the resulting formulae in a slightly different form by introducing $\langle d_q \rangle $ as the spatially constant inhaled quanta dose 
\begin{eqnarray}
  \langle d_q(t) \rangle & = & Q_{\mathrm{b}}\int_0^t \langle c(t')\rangle \; \mathrm{d}t' \label{eq:dose}  \\ 
  \langle d_q(t) \rangle & = & G_q\frac{Q_{\mathrm{b}}}{Q_{\mathrm{eff}}}
                \left(t + \frac{V}{Q_{\mathrm{eff}}}\left(\exp{\left(-\frac{Q_{\mathrm{eff}}}{V}t \right)} - 1\right) \right)   \label{eq:dose_explicit} \\
   \langle P(t)\rangle & = & 1 - \exp{\left(- \langle d_q(t) \rangle \right)}  \label{eq:Pgeneral} , 
\end{eqnarray}
where $Q_{\mathrm{b}}$ is the expected breathing volume flow rate of the susceptible persons. 

We generalize this theory for spatially variable situations in Appendix~\ref{app:P_extension} by replacing the spatial-averaging by ensemble averaging in the definition of the infection probability. This leads to reformulation of Eqn.~\eqref{eq:Pgeneral} as follows 
\begin{eqnarray}
  P(\xb,t) 
  & = & 1 - \exp{\left(-Q_{\mathrm{b}}\int_0^t \langle c^{(k)}(\xb,t') \rangle_e \; \mathrm{d}t'  \right)} \nonumber \\
  & = & 1 - \exp{\left(-\langle d_q^{(k)}(\xb,t) \rangle_e\right)},  \label{eq:Pgeneral3D}
\end{eqnarray}
where $k$ refers to an individual experiment and $\langle \cdot \rangle_e$ denotes ensemble averaging over all $N$ experiments in the ensemble, see Appendix~\ref{app:P_extension} for detailed description and derivation.

\begin{figure*}[ht]
    \centering
    \includegraphics[width=0.85\textwidth]{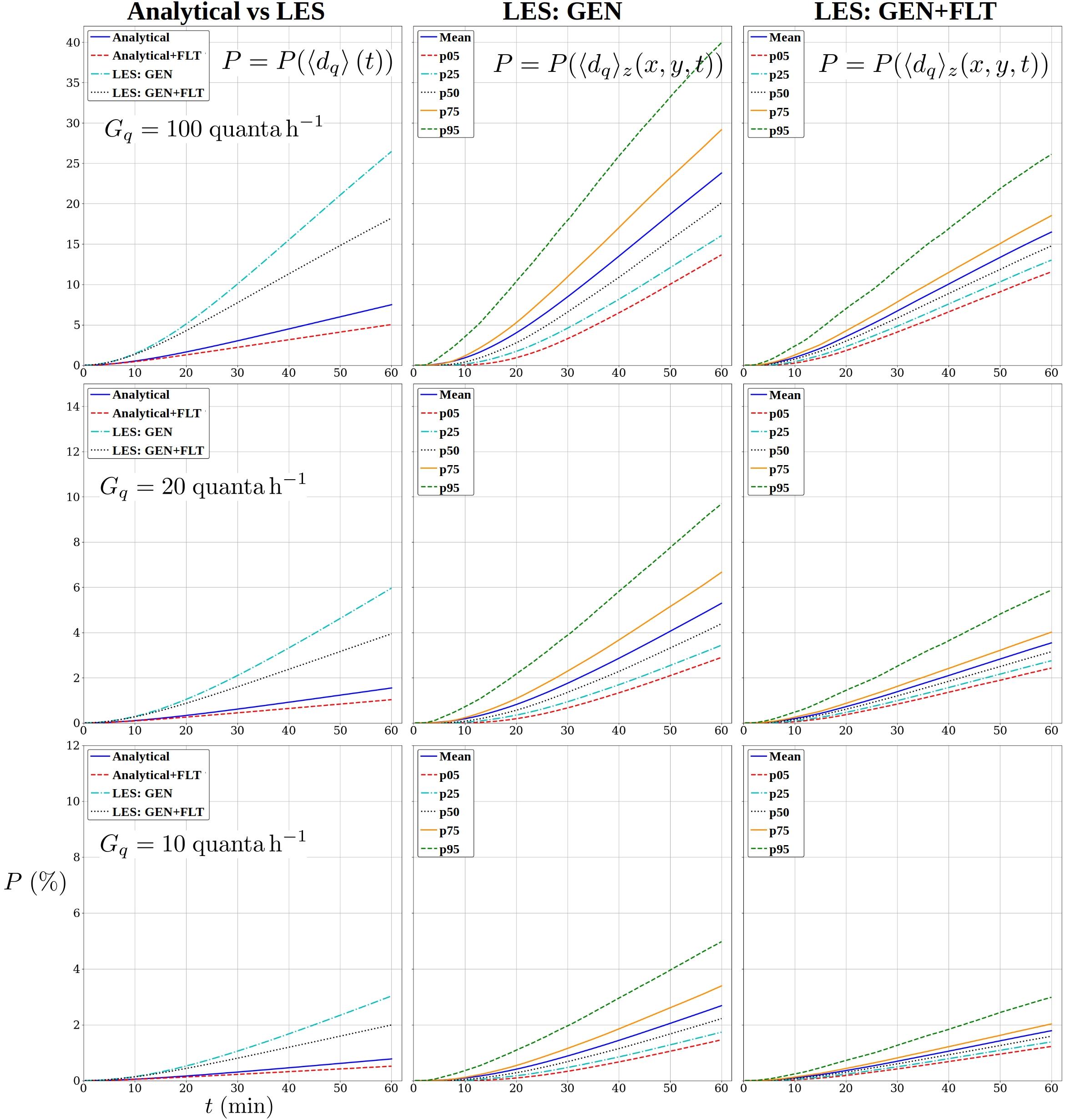}
    \caption{Infection probability as a function of time (s) since the arrival of the infecting person in the reference case. Left: comparison of Eqn.~\eqref{eq:Pgeneral3D} using the spatially averaged concentrations from LES with the fully analytic solutions~\eqref{eq:Pgeneral} using concentration from Eqn.~\eqref{eq:dose_explicit}. Middle: spatially varying probabilities according to Eqn.~\eqref{eq:Pgeneral3D} for the GEN case as  percentiles: 5th, 25th, 50th, 75th and 95th, and also the mean. Right: the same as in the middle, but for the GEN+FLT case.}
    \label{fig:Pref_comparisons}
\end{figure*}

In Fig.~\ref{fig:Pref_comparisons} we compare this approach with probability estimates obtained from Eqns.~\eqref{eq:dose_explicit} and~\eqref{eq:Pgeneral}. We evaluate the infection probabilities for three quanta generation rates $G_q$: 100, 20 and $\unit[10]{quanta~h^{-1}}$. This choice is motivated by the log-normal probability-density function (PDF) for $G_q$ proposed by \citet{Buonanno2020a}. According to this PDF, the most probable quanta rate is $\unit[20]{quanta~h^{-1}}$. We selected this value and the values $\unit[10]{quanta~h^{-1}}$ and $\unit[100]{quanta~h^{-1}}$ symmetrically to the most probable value $\unit[20]{quanta~h^{-1}}$ in terms of $\log{G_q}$, and corresponding roughly to 50\% probability. On the other hand,  \citet{Buonanno2020a} estimated their PDF based on data collected before the emergence of the delta-variant of SARS-CoV-2 which is observed to be significantly more infective than the earlier alpha-variant. Therefore, we assume that the highest selected quanta rate $\unit[100]{quanta~h^{-1}}$ might well be representative for the delta-variant in dining restaurant situations. It should be understood that the quanta-rate estimations vary considerably from case to case and all the three values selected here are probably rather conservative estimations. For example, singing may increase the quanta-emission rate significantly. \citet{Miller2020} estimated a quanta rate of a single infected singer in the Skagit Valley Chorale super-spreading event as high as $\unit[970\pm 390]{quanta~h^{-1}}$. Here, a constant inhalation rate of $\unit[800]{dm^3 \, h^{-1}}$ is assumed following e.g. \citet{Foster2021}. 

The left-hand column of Fig.~\ref{fig:Pref_comparisons} shows the time evolution of probabilities for the GEN- and GEN+FLT-cases based on LES-predicted concentrations averaged over the whole space, and compares these with the corresponding Wells-Riley predictions. The comparisons reveal the strong underestimation by the Wells-Riley model. These figures also show that the increase of $Q_{\text{eff}}$ by 65\% using the filtrating air purifiers decreases the probabilities by 31\% at $t=\unit[1]{h}$ while the asymptotic long-time relative reduction (relative change in $G_q/Q_\mathrm{eff}$) would be 39\%. The plots in the middle and right-hand columns quantify the spatial variability of the probability according to the LES results for the GEN and GEN+FLT cases respectively. These plots show the following percentiles: 5th, 25th, 50th (median), 75th and 95th, and also the mean for comparison. These probabilities are based on concentration fields averaged vertically over the so-called \textit{living zone} with $z$ ranging from $\unit[0.1]{m}$ to $\unit[2]{m}$. A vertical circular cylinder with a radius of $\unit[2]{m}$ centered around the source is excluded from the computation of the percentiles. The reason for this is that we focus on the longer-distance aerosol transmission omitting the immediate vicinity where direct droplet transmission may also occur, the infection risk is obviously very high and the aerosol concentrations are always highly specific to the nature of the respiratory activity.

\subsection{Risk-reduction strategies}

\begin{figure*}[ht]
    \centering
    \includegraphics[width=0.85\textwidth]{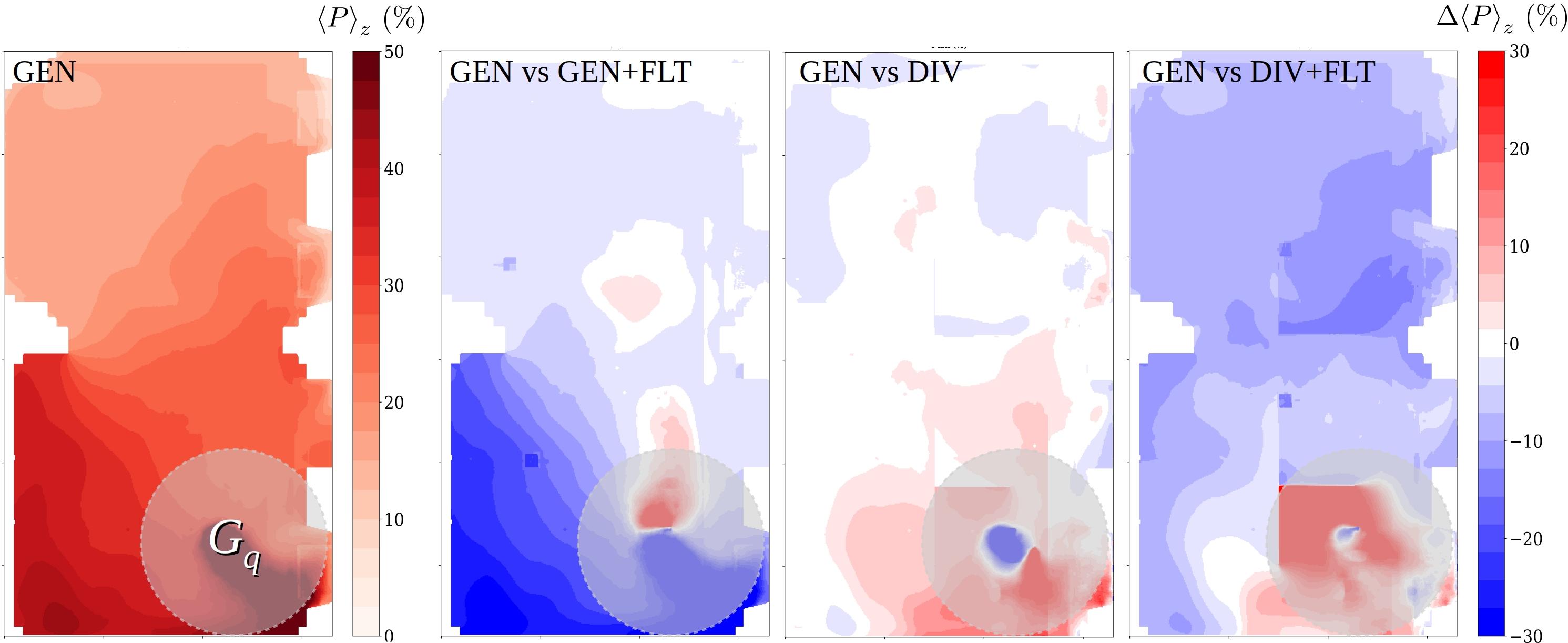}
    \caption{Effects of the risk-reduction strategies in infection-probability distributions over the room  at $t=\unit[1]{h}$ for $G_q = \unit[100]{h^{-1}}$. On the left (GEN) is the infection-probability distribution for the GEN case, and the other plots show the differences $\Delta \langle P \rangle_z$ due to the risk-reduction strategies for the cases GEN+FLT, DIV, and DIV+FLT in percentage units such that negative (blue) means reduction. The probabilities are vertically averaged over the living zone, i.e. the range $\unit[0.1]{m} \leq z  \leq \unit[2.0]{m}$. The $\unit[2]{m}$-radius-cylinder around the source denoting the near-source area, which is not of primary interest in this study, is shown as opaque grey disk in each plot.}
    \label{fig:P2d_comparison}
\end{figure*}

The effectiveness of the risk-reduction strategies are examined using Fig.~\ref{fig:P2d_comparison} which visualizes, as a reference, the probability distribution over the room for the GEN case at $t=\unit[1]{h}$ and, for comparison, the differences due to each risk-reduction strategy in percentage units. The quanta rate in this evaluation is $G_q = \unit[100]{h^{-1}}$. The probabilities are vertically averaged over the living zone, i.e. the range $\unit[0.1]{m} \leq z \leq \unit[2.0]{m}$. The difference plot GEN vs GEN+FLT, highlighting the effect of added air purifiers, shows a reasonable reduction in the probability as was already seen in Fig.~\ref{fig:Pref_comparisons}, and that the absolute reduction is largest within about $\unit[6]{m}$ from the source. However, the relative reduction is approximately equally important in the other end of the room. The difference plot GEN vs DIV shows that the space dividers provide no significant risk reduction as was hypothesized. The aerosols simply cannot be captured within semi-open compartments formed by this kind of space dividers, unless possibly in cases in which an air outlet vent happens to be located very near the compartment including the infection source. The difference plot GEN vs DIV+FLT shows again a reasonable risk reduction presumably due to the air purification. In this case the absolute reduction is more evenly distributed over the space than in the GEN+FLT case. This may be partially due to the space dividers, but probably also the placing of the air purifiers, which is different from the GEN+FLT case, plays a role here. 

In the light of these results, it is clear that none of the considered risk-reduction attempts is able to reduce the risk to a safe level in this case. Only a reasonable risk reduction is achieved by using the air purifiers at volume-flow rate which increases the effective ventilation rate by approximately 60\% or more. Drastic increase of the air-purifier volume-flow rate capacity would naturally lead to further risk reduction but the added cost and noise level and possible other disadvantages would probably make this option impractical.           

\section{Conclusions} \label{sec:conclusions}

A detailed LES study on indoor turbulence and its effect on dispersion of airborne respiratory pathogens under realistic ventilation conditions is reported. The focus lies on the long-distance aerosol transmission and relevant flow mechanisms governing aerosol dispersion indoors. 

The study involves an aerosol particle dispersion experiment and a LES model validation against the measured concentrations in 27 locations inside the restaurant room. The level of agreement between the experimental and LES modelled concentration time series is established with the set of performance metrics adopting a strict set of evaluation criteria used in air-quality modeling \citep{Chang2004,Moonen2013}.  The level of detail and breadth of the validation enterprise is deemed high enough to provide sufficient evidence that the documented LES model is capable of resolving the relevant flow physics governing the spread of pathogen-carrying aerosols indoors.     

The role of turbulent mixing in the context of continuous emission event is demonstrated to be an important mechanisms in controlling indoor hygiene. Higher turbulence level dilutes and disperses local concentration peaks faster, consequently leading to faster activation of ventilation outlets and added filtration in exhausting pathogens from the indoor space.

The aerosol concentration results are further refined to obtain estimates for the infection probabilities. The Wells-Riley model is formally and rigorously extended to rely on CFD-predicted time- and space-dependent realistic concentration fields and to yield time- and space-dependent realistic infection probability fields.

In the infection-risk analysis, the original Wells-Riley model gave three to four times smaller mean infection probabilities than the LES-based calculation. This underestimation is due to the perfect and instantaneous mixing assumption embedded in the analytical model, which leads to the omission of local peak concentrations and overestimation of the ventilation exit term. The underestimation stems from the mean concentration, subsequently affecting the inhaled cumulative doses and finally infection probabilities. The extent of underestimation varies depending on the relative positions of the infection source(s) and the ventilation outlet vent(s). The two air purifier units used in the study increased the existing ventilation capacity by 65\%. This leads to a relative reduction of 39\% in the nearly asymptotic ($\unit[1]{h}$) mean risk probability according to the original Wells-Riley model, whereas the LES model yielded about 31\% for the relative reduction. 

Two risk-reduction strategies are assessed using the LES-based infection probability modeling. Both, spatial distribution and mean evolution of the infection risk are examined.  With balanced placement of the air purifier units in the generic configuration, the risk reduction is most prominent close to the source where concentrations are the highest. Due to the spatial distribution of the infection risk, the analytical model grossly underestimate the level of infection risk and its evolution within real spaces, but they also underestimate the local efficiency of air purifiers. However, dramatic reduction in the overall infection risk level requires a high relative capacity of air purifiers (compared to the existing ventilation capacity) may, therefore, be very  expensive and impractical solution. The second strategy which relies on space partitioning is confirmed not to reduce the infection risk. The partitioning does lead local accumulation of infection-risk, but the general risk level is not reduced. Therefore, it should not be considered a risk-reduction strategy alone. When air purifiers are applied together with space dividers, the situation improves dramatically. But this approach requires that added filtration units should be added to each individual segment. Otherwise the risk distribution may become highly variable. Thus, best efficiency from a small number of air purifier units is obtained without obstructing the air flow within the indoor space.

\section*{Authors' contributions}
M.A. and A.H. conceptualized the computational study and designed the computational model set up. A.H. and M.S. implemented the necessary modifications in the PALM model code. M.A. built the computational model set up and performed the simulations and the data post-processing. J.K. designed and constructed the sensor network system. J.K. and T.G. designed the experimental set up and performed the measurements and data analysis. A.H. derived the extension of the Wells-Riley model for spatially variable concentration fields. All authors contributed to writing the manuscript.

\begin{acknowledgments}
This study was financially supported by Business Finland Corona Co-Creation funding 40\,988~/ 31~/ 2020 and by Academy of Finland COVID-19 special funding grant number 335\,681. We warmly thank restaurateur chef Henri Alen for kindly providing one of his restaurants for our use to carry out the experiments, and UniqAir~Ltd for providing the air purifiers for the experiments.
\end{acknowledgments}

\section*{Data Availability Statement}

The PALM model system is freely available at \url{http://palm-model.org} and distributed under the GNU General Public Licence v3 (\url{http://www.gnu.org/copyleft/gpl.html}). However, the simulations presented in this article were performed using a modified code based on PALM revision 4786. This modified source code (4786M) is available at \url{https://doi.org/10.5281/zenodo.5596111} \citep{r4786M2021}. The data supporting the findings and conclusions of this study are available within the article. More extensive background data such as the room geometry is available from the corresponding author upon reasonable request.

\clearpage
\appendix
\section{Model evaluation results}
\label{app:evaluation_results}
Validation comparisons for the window-side mast row are shown in Figs.~\ref{fig:expr_les_genW} and~\ref{fig:expr_les_fltW}. The comparisons for the center- and door-side mast row are presented here.
\begin{figure*}[hb!]
    \centering
    \includegraphics[width=0.8\textwidth,trim={0.0cm 0.0cm 0.0cm 0.0cm},clip]{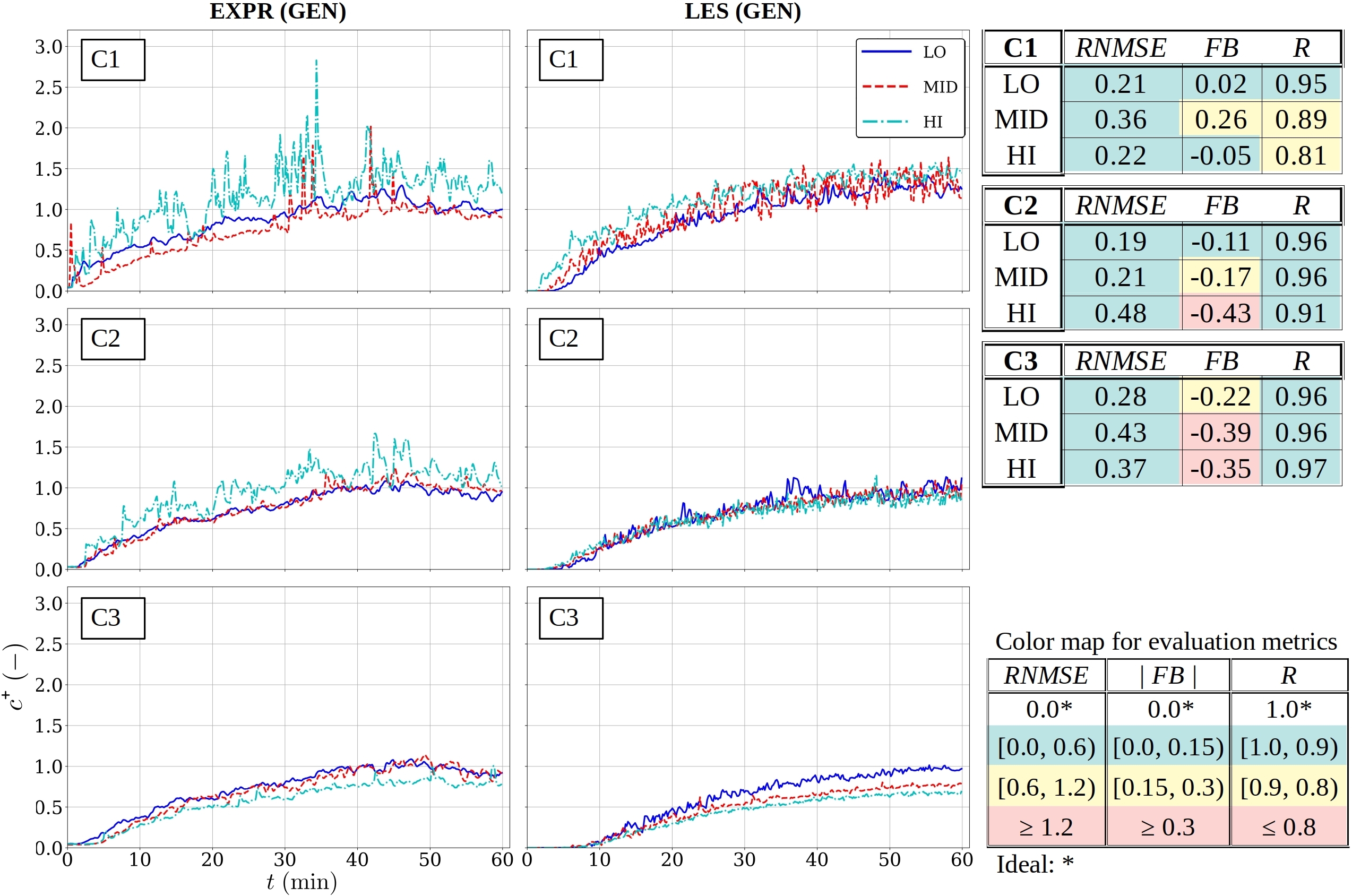} 
    \caption{Comparison of measured (left) and modelled (middle) normalized concentration time series in the center mast row without the air purifiers (GEN) and tabulated evaluation metrics (right) with color coding. The color coding and the acceptance criteria are tabulated on the lower right corner.}
    \label{fig:expr_les_genC}
\end{figure*}

\begin{figure*}[hb]
    \centering
    \includegraphics[width=0.8\textwidth,trim={0.0cm 0.0cm 0.0cm 0.0cm},clip]{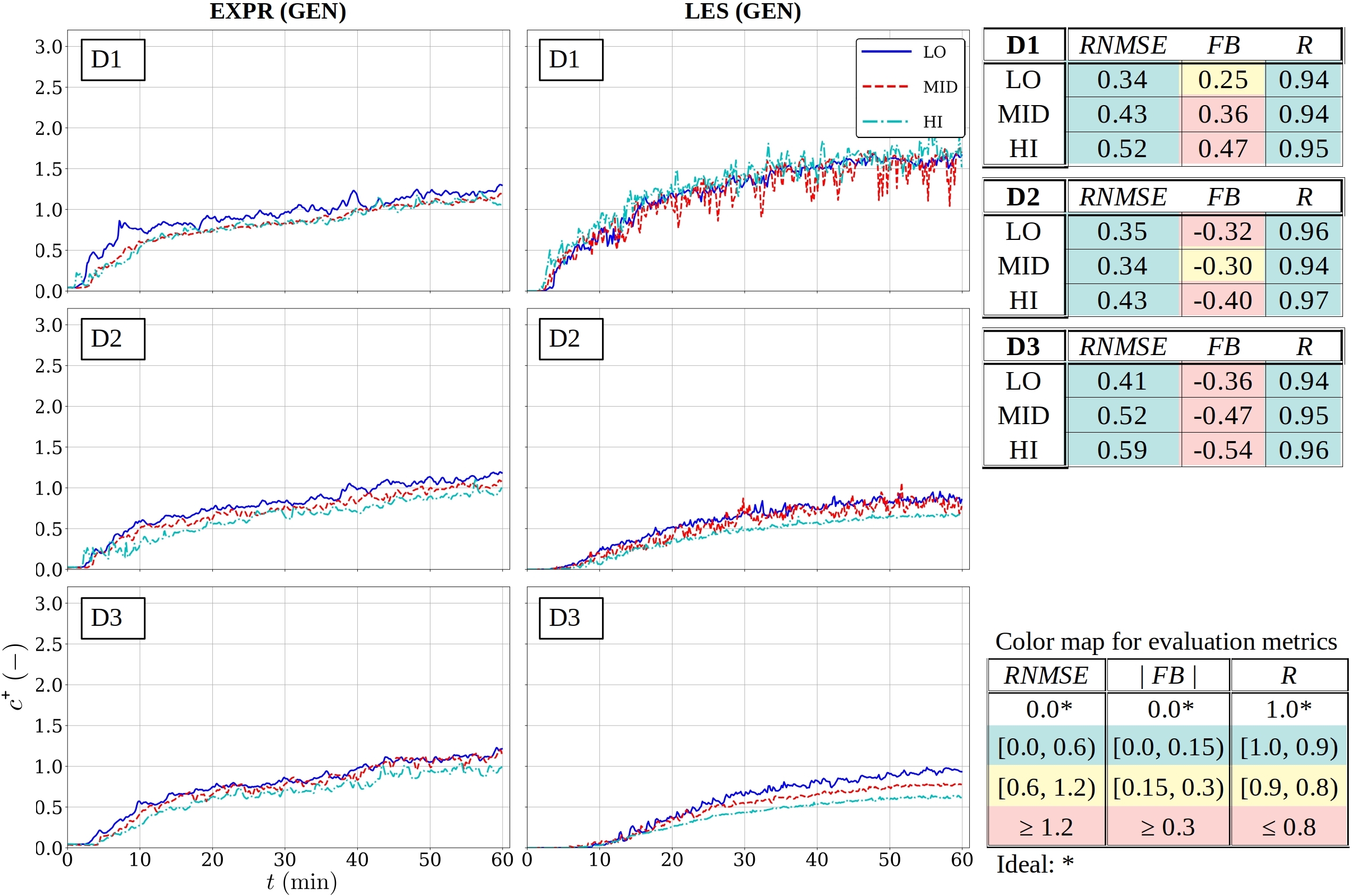}
    \caption{The same as in Fig.~\ref{fig:expr_les_genC} but for the door-side mast row.}
    \label{fig:expr_les_genD}
\end{figure*}

\begin{figure*}[t]
    \centering
    \includegraphics[width=0.8\textwidth,trim={0.0cm 0.0cm 0.0cm 0.0cm},clip]{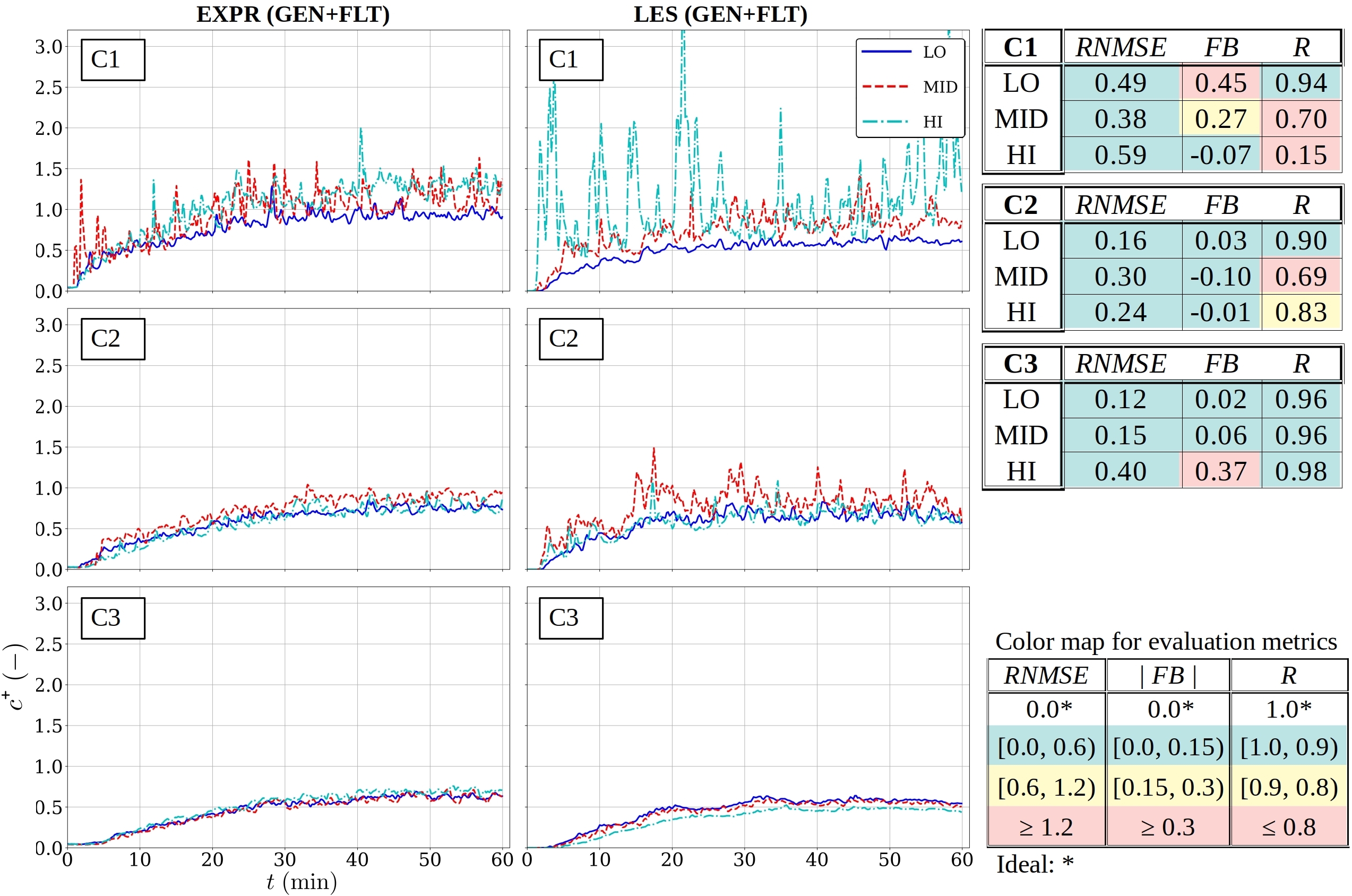}
    \caption{The same as in Fig.~\ref{fig:expr_les_genC} (center mast row) but  with the air purifiers (GEN+FLT).}
    \label{fig:expr_les_fltC}
\end{figure*}

\begin{figure*}[b]
    \centering
    \includegraphics[width=0.8\textwidth,trim={0.0cm 0.0cm 0.0cm 0.0cm},clip]{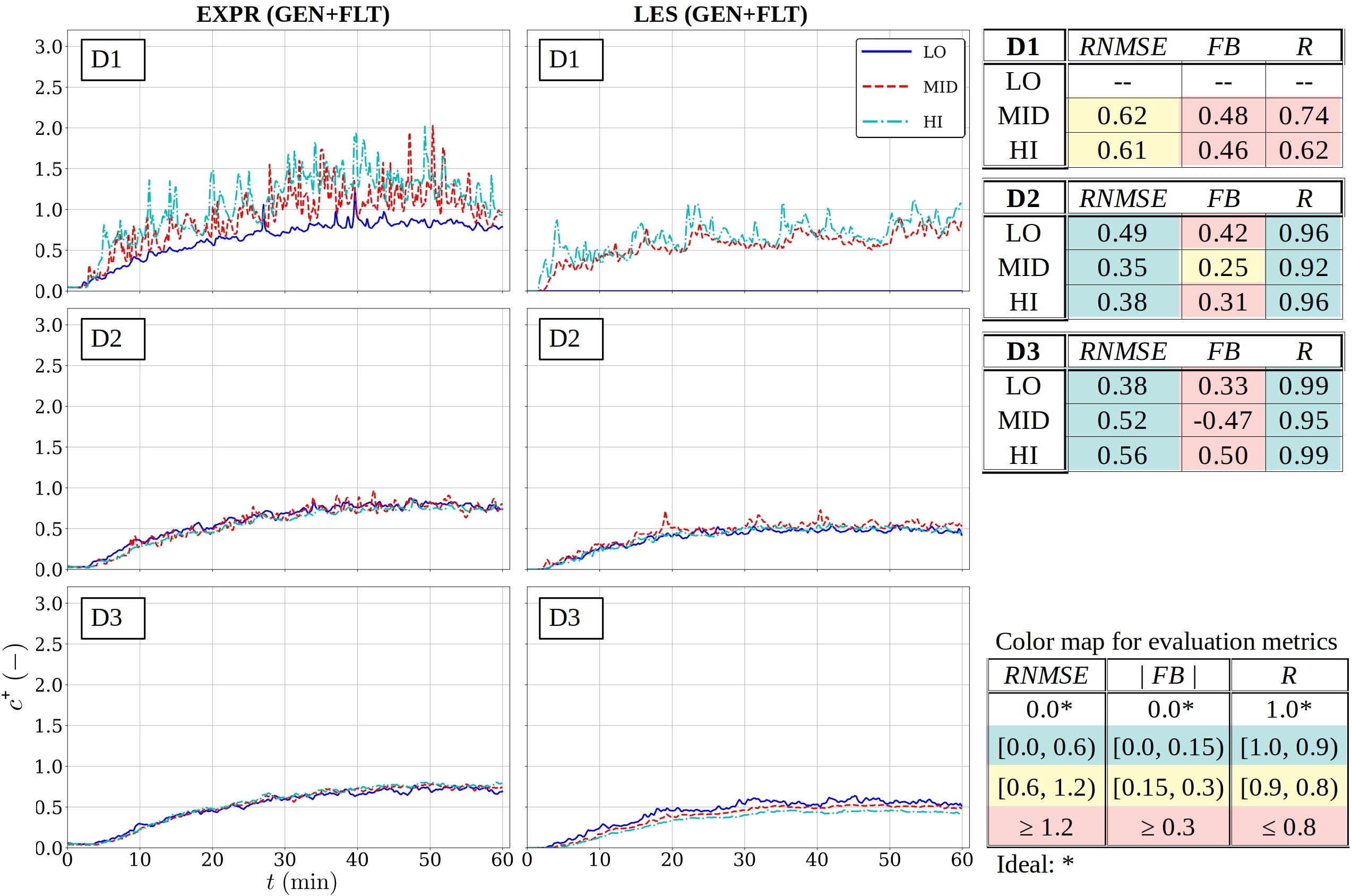}
    \caption{The same as in Fig.~\ref{fig:expr_les_genC} but  for the door-side mast row and with the air purifiers (GEN+FLT).}
    \label{fig:expr_les_fltD}
\end{figure*}

\FloatBarrier

\section{Extension of the infection probability model to spatially inhomogeneous situations}
\label{app:P_extension}

The time dependent probability of infection $\langle P(t)\rangle$ is derived by Gammaitoni and Nucci \cite{Gammaitoni1997b}. 
It is based on the analytical solution of the one-way coupled differential equation pair for the spatially constant quanta concentration $c(t)$ and the number of susceptible persons $S(t)$ 
\begin{eqnarray} 
    && \frac{\mathrm{d}c}{\mathrm{d}t} 
       = \frac{1}{V}\left( G_q - Q_{\mathrm{eff}}c \right)     \label{eq:c_hom} \\
    && \frac{\mathrm{d} S}{\mathrm{d}t} 
       = -Q_b c S                                              \label{eq:S_hom} \\
    && c(0) = 0                          \nonumber \\
    && S(0) = S_0 \geq 0.                \nonumber
\end{eqnarray}
Here $V$ is the volume of the indoor space, $G_q$ is the quanta generation rate (quanta/s), $Q_{\mathrm{eff}}$ is the effective air exchange rate including the volume-flow rate through the air purifiers if present (here we assume the purification efficiency is 100\%), and 
$Q_{\mathrm{b}}$ is the expected constant breathing volume-flow rate of the susceptible persons. In this study we assume zero initial concentration although this assumption is not necessary. 
\citet{Gammaitoni1997,Gammaitoni1997b} 
assumed that $S_0$ non-infected susceptible persons are inside the indoor space in question together with an infective person. As time elapses, $S$ will reduce as they gradually get infected. Their location and possible movements within the space do not matter due to the assumption of uniform distribution of quanta concentration, which is necessary if concentration is to be obtained by solving Eqn.~\eqref{eq:c_hom}.

The solutions for $c(t)$ and $S(t)$ are
\begin{eqnarray}
  c(t) & = & \frac{G_q}{Q_{\mathrm{eff}}}
         \left(1 - \exp{\left(-\frac{Q_{\mathrm{eff}}}{V}t \right)} \right) \label{eq:c_hom_soln}  \\ 
  S(t) & = & S_0 \exp{\left(-Q_{\mathrm{b}}\int_0^t c(t') \mathrm{d}t'\right)} \nonumber \\
       & = & S_0 \exp{\left(-d_q(t)\right)} \label{eq:aS_hom}
\end{eqnarray}
with 
\begin{equation}
    d_q(t)= G_q\frac{Q_{\mathrm{b}}}{Q_{\mathrm{eff}}}
            \left(t + \frac{V}{Q_{\mathrm{eff}}}\left(\exp{\left(-\frac{Q_{\mathrm{eff}}}{V}t \right)} - 1\right) \right), \label{eq:dq_hom}
\end{equation}
where, $d_q(t)$ is spatially constant cumulative inhaled quanta dose. The infection probability is then identified as the number of infected divided by the initial number of susceptibles as 
\begin{equation}
  P(t) = 1 - S(t)/S_0 = 1 - \exp{\left(- d_q(t) \right)}  \label{eq:P_WR} 
\end{equation}

However, concentration fields in real indoor spaces can seldom been considered constant in space. If we wish to employ spatially variable concentration-field information for instance from our LES instead of Eqn.~\eqref{eq:c_hom_soln}, 
we need to reformulate the interpretation of the infection probability as described by the following thought experiment. Let us assume we have an experiment with an arbitrary number of susceptible persons seated still in different labeled locations $\xb$ within the room together with one or more infective persons in known locations. This experiment is repeated with identical setting having a new ensemble of equal number of non-infected susceptibles. In total, the experiment is repeated $N \gg 1$ times keeping everything else unchanged except the people and the turbulent flow and concentration field realization which vary from experiment to experiment although in the statistical sense the fields in all $N$ experiments are identical. Now, the infection probability at each location is redefined as 
\begin{equation}
  P(\xb,t) = 1 - \langle S^{(k)}(\xb,t) \rangle_e,  \label{eq:Pgen3D}
\end{equation}
where $k$ refers to an individual experiment and $\langle \cdot \rangle_e$ denotes ensemble averaging over all $N$ experiments. This exchange of interpretation is justified based on the ergodic hypothesis. The equation~\eqref{eq:S_hom} 
is now rewritten as
\begin{eqnarray}
  && \frac{\mathrm{d} \langle S^{(k)} \rangle_e}{\mathrm{d}t} = -Q_b \langle c^{(k)} \rangle_e \langle S^{(k)}\rangle_e \label{eq:S_inhom} \\
  && \langle S^{(k)}(\xb,0) \rangle_e  = 1 \label{eq:Sk_t0}
\end{eqnarray}
Also $S^{(k)}(\xb,0) = 1$ for all $k$. There is no equation for concentration, instead, spatially and temporally variable concentration field has to be obtained separately from a CFD-analysis. Note that $S^{(k)}$ of an individual experiment $k$ in our thought experiment can only obtain the integer values zero or one, but the actual variable to be solved $0 \leq \langle S^{(k)}(\xb,t) \rangle_e \leq 1$ is continuous and real-valued.  The variables $S$ and $c$ are now functions of space, but there is no spatial coupling in Eqn.~\eqref{eq:S_inhom}. 
Therefore it can be solved in any point $\xb$ independently of any other points. The solution is mathematically similar to Eqn.~\eqref{eq:aS_hom}, 
but it is rewritten here using appropriate notation as
\begin{eqnarray}
  \langle S^{(k)}(\xb,t) \rangle_e 
    &=& \exp{\left(-Q_{\mathrm{b}}\int_0^t \langle c^{(k)}(\xb,t') \rangle_e \, \mathrm{d}t'\right)} \nonumber \\
    &=& \exp{\left(-\langle d_q^{(k)}(\xb,t) \rangle_e\right)} \label{eq:Sk}.
\end{eqnarray}
Thus, Eqn.~\eqref{eq:Pgen3D} becomes
\begin{equation}\label{eq:Pgen3D_explicit}
  P(\xb,t) = 1 - \exp{\left(-\langle d_q^{(k)}(\xb,t) \rangle_e\right)}.
\end{equation}
%
%

\bibliography{bibliography}

\end{document}